\documentclass[]{jfm}
\usepackage[applemac]{inputenc}
\usepackage[english]{babel}
\usepackage{listings} 
\usepackage{newtxtext}
\usepackage{newtxmath}
\usepackage{curve2e}
\usepackage{fancyhdr}
\usepackage{lscape}
\usepackage{graphicx}
\usepackage[caption=false]{subfig}
\usepackage{mathtools}
\usepackage{amsfonts}
\usepackage{acronym}
\usepackage{psfrag}
\usepackage{amssymb}
\usepackage{ifsym}
\usepackage{nomencl}
\usepackage{multirow}
\usepackage{multicol}
\usepackage{placeins}
\usepackage{color}
\usepackage{cases}
\usepackage[standard]{frontespizio}
\usepackage{float}
\usepackage{subfloat}
\usepackage[breaklinks=true]{hyperref}
\hypersetup{
    colorlinks = true,
    urlcolor   = blue,
    citecolor  = black,
}

\newcommand{\RomanNumeralCaps}[1]
\linenumbers
\usepackage{enumerate}
\usepackage{rotating}
\usepackage{rotfloat}
\usepackage{tabulary}
\usepackage{natbib}
\usepackage{stmaryrd}
\usepackage{setspace}
\usepackage{verbatim}
\usepackage{breakcites}
\usepackage{ar}
\usepackage{tikz,tikz-3dplot}
\usepackage{url}

\usetikzlibrary{arrows,shapes,calc}
\usetikzlibrary{decorations.pathreplacing}

\newcommand\St{\mbox{\textit{St}}}  
\newcommand\Nu{\mbox{\textit{Nu}}} 

\definecolor{darkviolet}{rgb}{0.58, 0.0, 0.83}
\definecolor{shamrockgreen}{rgb}{0.0, 0.62, 0.38}
\definecolor{deepskyblue}{rgb}{0.0, 0.75, 1.0}
\definecolor{amber}{rgb}{1.0, 0.49, 0.0}
\definecolor{red}{rgb}{1.0, 0.0, 0.0}

\newcommand\diff{\mbox{\textit{d}}}  

\title{Direct numerical simulation of forced thermal convection in square ducts up to $\Rey_{\tau} \approx 2000$}
\author{Davide Modesti$^{1}$\corresp{\email{d.modesti@tudelft.nl}} \and Sergio Pirozzoli$^2$}

\affiliation{\aff{1} Faculty of Aerospace Engineering, Delft University of Technology, Kluyverweg 2, 2629 HS Delft, The Netherlands
\aff{2} Dipartimento di Ingegneria Meccanica e Aerospaziale, Sapienza Universit\`a di Roma, via Eudossiana 18, 00184 Roma, Italia}

\begin{document}
\maketitle

\begin{abstract}
We carry out direct numerical simulation (DNS) of flow in a turbulent square duct 
by focusing on heat transfer effects, considering the case of unit Prandtl number. 
Reynolds numbers up to $\Rey_{\tau} \approx 2000$ are
considered which are much higher than in previous studies, and 
which yield clear scale separation between inner- and outer-layer dynamics.
Close similarity between the behavior of the temperature and the streamwise velocity fields 
is confirmed as in previous studies related to plane channels and pipes. 
Just like the mean velocity, the mean temperature is found to exhibit 
logarithmic layers as a function of the nearest wall, however with a different
slope. The most important practical implication is the validity of 
the traditional hydraulic diameter as the correct reference length for reporting
heat transfer data, as we rigorously show here. 
Temperature and velocity fluctuations also have similar behavior, 
but apparently logarithmic growth of their inner-scaled peak variances is not 
observed here unlike in canonical wall-bounded flows. Analysis of the split contributions to 
the heat transfer coefficient shows that mean cross-stream convection associated
with secondary motions is responsible for about $5\%$ of the total. 
Finally, we use the DNS database to highlight shortcomings of traditional linear closures for the
turbulent heat flux, and show that substantial modeling improvement may be in principle obtained 
by retaining at least the three terms in the vector polynomial integrity basis expansion.
\end{abstract}

\begin{keywords}
\end{keywords}

\section{Introduction}

Heat transfer in internal flows is a subject of utmost relevance in mechanical and aerospace engineering applications.
A large amount of experimental and numerical studies have been carried out in the past, and 
a variety of analytical and semi-empirical prediction tools have been developed, 
which are extensively reported in classical textbooks~\citep{kays_93,rohsenow_98}. 
Most studies have been carried out for the canonical case of 
ducts with circular cross section or planar channels, whereas much less is known
about the case of ducts with more complex geometry, which
also have great practical relevance, for instance in water draining or ventilation systems, 
nuclear reactors, heat exchangers, space rockets and turbomachinery.
In that case, the typical engineering approach is to use the same
correlations established for the case of circular pipes, by replacing the
pipe diameter with the hydraulic diameter of the duct~\citep{kays_93,white_06}.
Although this approach is found to be rather successful in practice, it lacks 
solid theoretical foundations. Furthermore, the large scatter in experimental
data makes it difficult to quantify the actual accuracy of semi-empirical 
prediction formulas, which are reported to have $\pm 9\%$ uncertainty 
for smooth ducts with uniform heating~\citep{shah_98}.

Heat transfer in square ducts was first studied experimentally by \citet{brundrett_67}, who considered 
air flow at bulk Reynolds number $\Rey_b = H u_b/\nu$ (where $H=2h$ is duct side length and also duct hydraulic diameter, $u_b$ is the bulk velocity, 
and $\nu$ the fluid kinematic viscosity) 33000 and 67000. Close similarity between the wall shear stress
and the heat flux distributions was shown (with quoted discrepancy of $\pm 2\%$), 
which the authors connected with similar mixing action of the
secondary currents on momentum and temperature. As a result, they observed that the ratio of the
average friction and heat-transfer coefficients for a square duct is approximately the same as
for a circular pipe. Measurements of the wall-normal temperature profiles highlighted 
close universality when the local wall heat flux is used for normalization, and the presence 
of a sizeable logarithmic layer, with extrapolated value of the scalar K\'arm\'an constant
of $\kappa_{\theta} \approx 0.51$.
Those findings were qualitatively supported from later experiments by \citet{hirota_97}, who also analysed
temperature fluctuations and velocity/temperature fluctuations correlations, and found significant 
distortions over the cross section associated with the secondary motions. In that 
study, the inferred scalar von K\'arm\'an constant was $\kappa_{\theta} \approx 0.46$,
hence more similar to the values generally quoted for circular pipe flow, namely 
$\kappa_{\theta} \approx 0.47$~\citep{kader_72}.

Early computational studies of heat transfer in square ducts mostly relied on the use of RANS models. 
For instance, \citet{launder_73} developed a full Reynolds stress closure 
to predict flow and heat transfer in a square duct. Although 
the global heat transfer coefficient showed general underprediction by about $10\%$,
the distribution of the wall heat flux qualitatively reproduced the experimental data of \citet{brundrett_67}.
High-fidelity computational studies of heat transfer in square ducts have been quite limited so far.
\citet{vazquez_02} first studied turbulent flow
through a heated square duct by means of large-eddy simulation (LES) at moderate Reynolds number ($\Rey_b=6000$),
considering the case in which one of the walls is hotter than the other three.
Accounting for fluid viscosity variation with temperature, they found that
turbulent structures near the hot wall become larger than near the other walls,
in such a way that wall scaling is satisfied.
LES with different wall thermal states, also in the presence of duct rotation, were carried out by
\citet{pallares_02}. 
\citet{yang_09} carried out under-resolved DNS (direct numerical simulation) of turbulent flow in a square duct with 
uniform volumetric heating, finding good agreement with the temperature profiles 
measured by \citet{hirota_97}, and with predictions of the heat transfer coefficient 
resulting from Gnielinski's analogy~\citep{gnielinski_75}.
Large-eddy simulations of space-developing turbulent flow through a heated square duct were carried out by
\citet{schindler_19}, at bulk Reynolds number up to $\Rey_b = 10^4$. 
As noted by \citet{brundrett_67}, strong correlation of the local wall heat flux 
with the wall shear stress was found. 
~\citet{sekimoto_11} carried out DNS of mixed convection in square duct flow ranging from very low
to unit Richardson number. They found that modifications of the secondary flows due to buoyancy
start at Richardson number $\approx 0.025$, and the effect of buoyancy becomes dominant
at Richardson $\approx0.25$ and the eight cross-stream vortices are replaced by two large-scale ones.
Turbulence modelling of forced and natural thermal convection is a topic of primary interest for industrial applications,
but is has received less attention than its momentum counterpart.
Modelling approaches for the turbulent heat flux vector were reviewed by~\citet{hanjalic_02},
who pointed out inadequacy of the turbulent Prandtl number concept in complex flows, and advocated the use of second-moment closures
or algebraic models for turbulent heat transfer. However, the state-of-the art for modelling turbulent convection
has not advanced much since the time of that review article, and linear eddy-diffusivity models relying on use of the turbulent Prandtl number are routinely used.

In this paper we study heat transfer in fully developed square duct flow with uniform internal heating
and isothermal walls, by carrying out DNS at much higher Reynolds number
than in previous computational studies. Although relatively simple, this setup includes the main complicating effects
involved with non-trivial cross-sectional geometries, and primarily associated with the
formation of secondary motions of the second kind~\citep{prandtl_27,nikuradse_30}.
The assumption herein made of unit molecular Prandtl number (defined as the ratio of the kinematic viscosity
to the scalar diffusivity, $\Pran=\nu/\alpha$), is instrumental to more closely scrutinizing differences 
between the behavior of the streamwise velocity field and of passively advected scalars.
The present study is the continuation of previous efforts~\citep{pirozzoli_18,modesti_18,orlandi_18}
targeted to studying turbulent flows in square ducts by means of DNS.
In those studies we found that the intensity of the secondary motions is in the order of a few percent
of the duct bulk velocity, and despite their persistence they do contribute to the mean duct friction
by at most a few percent. 
In this study we aim to extend our analysis to the temperature field and
to quantify the effect of secondary flows on the wall heat flux.
Additionally, we assess the accuracy of the turbulent Prandtl number concept
and of constitutive relations for the turbulent heat flux vector.

\section{Methodology}\label{sec:method}
\begin{table}
\footnotesize
\centering
\begin{tabular}{lcccccccccccccc}
\hline
\hline
 Case & $\Rey_b$ & $\Rey_{\tau}^*$ & $C_f\times10^{3}$ & $\Nu$ & $N_x$ & $N_y$ & $N_z$ & $\Delta x^*$ & $\Delta z^*$ & $\Delta y_w^*$ & Symbol \\
\hline
 A & $4410$  & $150$  & $9.26$ & $ 18.1$ & $512$  &  $128$ &  $128$ & $5.6$ & $3.0$ & $0.55$ & \color{gray}$\bigcirc$ \\
 B & $7000$  & $227$  & $8.41$ & $ 26.9$ & $640$  &  $144$ &  $144$ & $6.6$ & $4.8$ & $0.51$ & \color{red}$\bigtriangledown$ \\
 C & $12000$ & $365$  & $7.40$ & $ 41.9$ & $768$  &  $208$ &  $208$ & $9.0$ & $5.3$ & $0.69$ & \color{cyan}$\bigtriangleup$ \\
 D & $17800$ & $519$  & $6.80$ & $ 57.6$ & $1024$ &  $256$ &  $256$ & $9.5$ & $6.3$ & $0.53$ & \color{shamrockgreen}$\triangleright$ \\
 E & $40000$ & $1055$ & $5.57$ & $109.0$ & $2048$ &  $512$ &  $512$ & $9.6$ & $6.4$ & $0.60$ & \color{amber}$\square$ \\
 F & $84000$ & $2041$ & $4.72$ & $197.0$ & $4096$ & $1024$ & $1024$ & $9.3$ & $6.0$ & $0.67$ & $\Diamond$ \\                   
\hline
 A0& $4410$  & $154$ & $9.66$ & $21.1$ & $512$  &  $128$ &  $128$ & $5.7$ & $3.3$ & $0.54$ &  -\\ 
 B0& $7000$  & $231$ & $8.58$ & $30.0$ & $640$  &  $144$ &  $144$ & $6.8$ & $4.7$ & $0.67$ &  -\\ 
 C0& $12000$ & $367$ & $7.48$ & $44.5$ & $768$  &  $204$ &  $204$ & $9.0$ & $5.4$ & $0.71$ &  -\\ 
 D0& $17800$ & $515$ & $6.62$ & $59.6$ & $1024$ &  $256$ &  $256$ & $9.5$ & $6.0$ & $0.79$ &  -\\ 
\end{tabular}
\caption{Flow parameters for square duct DNS.
Box dimensions are $6\pi h \times 2h \times 2h$ for all flow cases.
Flow cases denoted with the 0 suffix are carried out by suppressing the secondary motions.
$\Rey_b = 2 h u_b / \nu$ is the bulk Reynolds number, and
$\Rey_{\tau}^* = h u_{\tau}^* / \nu$ is the friction Reynolds number.
$C_f = 2 \tau_w /(\rho u_b^2)$ is the friction coefficient,
and $\Nu$ is the Nusselt number.
$\Delta x$ is the mesh spacing in the streamwise direction, and
$\Delta z$, $\Delta y_w$ are the maximum and minimum mesh spacings
in the cross-stream direction, all given in global wall units, $\delta_v^*=\nu/u_{\tau}^*$.}
\label{tab:test}
\end{table}

The numerical simulation of incompressible turbulent flow in non-circular ducts 
is a more challenging task for numerical algorithms than the canonical cases of plane channel and pipe flow.
The main reason resides in the availability of only one direction of space homogeneity,
which prevents the use of efficient inversion procedures for Poisson equations
based on double trigonometric expansions~\citep{kim_85,orlandi_12}.
This difficulty has been tackled using two-dimensional Poisson
solvers based on cyclic reduction~\citep{gavrilakis_92}, using algebraic multigrid methods~\citep{vinuesa_14,marin_16},
or fast diagonalization~\citep{pinelli_10},
resulting in a larger computational cost than channel and pipe flow simulations.
In the present work we use a fourth-order co-located finite-difference solver, previously
used for DNS of compressible turbulence, also in the low-Mach-number regime~\citep{pirozzoli_13,modesti_16,bernardini_21}.
Here, the convective terms in the Navier--Stokes equations are preliminarily expanded to
quasi-skew-symmetric form, in such a way to discretely
preserve total kinetic energy from convection~\citep{pirozzoli_10}.
Semi-implicit time stepping is used for time advancement in order to relax the acoustic time step limitation,
thus allowing efficient operation at low Mach number, also
through the use of the entropy evolution equation rather than the total energy equation~\citep{modesti_18a}.
The streamwise momentum equation is forced in such a way as to maintain a constant mass flow rate
(the spatially uniform driving term is hereafter referred to as ${\Pi}$),
periodicity is exploited in the streamwise direction, and isothermal no-slip boundary
conditions are used at the channel walls.
Let $h$ be the duct half-side, the DNS
have been carried out for a duct with $[-h:h] \times [-h:h]$ cross section,
and length $6 \pi h$. 

Six DNS have been carried out at bulk Mach number $M_b=u_b/c_w=0.2$ (where $c_w$ is the speed of sound at
the wall temperature), and bulk Reynolds number $\Rey_b=4400-84000$ (table~\ref{tab:test}), 
and hereafter labeled with letters from A to F. 
In order to quantify the effect of secondary flows we have
also carried out simulations at the same bulk Reynolds numbers, wherein secondary flows
are numerically suppressed, and which are denoted with the suffix 0.
The turbulence Mach number $M_t=u'/{c_w}$ nowhere exceeds $0.01$ for any of the simulations,
hence the present DNS may be regarded as representative of genuinely incompressible turbulence.
Issues related to size of the computational box, mesh resolution and statistical convergence
were discussed in a previous publication~\citep{pirozzoli_18}, which focused on the analysis
of the velocity statistics. 
In the present study the Navier--Stokes equations are augmented with the transport equation for a 
passive scalar field, with molecular diffusivity $\alpha = \nu$, in such a way that the
molecular Prandtl number is unity for all simulations. Although the study of passive scalars
may be relevant in several fluid dynamics applications, from now on we will always refer to the 
transported scalar as representative of the temperature field, and the associated 
fluxes as representative of heat fluxes.
Similarly to the streamwise velocity field, the passive scalar equation is also forced with a 
time-varying, spatially homogeneous forcing term (say $Q$), in such a way that its mean value 
is maintained in time~\citep{pirozzoli_16}. 
This forcing method corresponds to adding uniform bulk heating
to balance heat losses, although other forcing strategies are possible~\citep[e.g.][]{piller_05,abe_17}.
The most common alternative to uniform bulk heating is enforcement of constant heat flux in time. Both approaches have advantages
and disadvantages.
Uniform bulk heating is less realistic as it is difficult to attain in experiments~\citep{piller_05}, however
it is more efficient from a computational point of view because of faster convergence towards a statistically stationary state.
In the present work we opt for uniform bulk heating in light of higher computational efficiency in DNS of high-Reynolds-number flows.
We expect that this approach may yield slightly higher heat flux number as compared to the case of strictly constant heat flux~\citep{abe_17,alcantara_21}.
In the following 
we use a different notation for the mean temperature in the duct,
\begin{equation}
	\theta_b = \frac 1{A_c} \int_{A_c} (\Theta-\theta_w) \diff A_c , \label{eq:thetab}
\end{equation}
and for the bulk temperature, 
\begin{equation}
	\theta_m = \frac 1{u_b A_c} \int_{A_c} U (\Theta-\theta_w) \diff A_c , \quad u_b = \frac 1A_c \int_{A_c} U \diff A_c , \label{eq:thetam}
\end{equation}
where $\theta_w$ is the wall temperature, and $A_c$ is the cross-sectional area of the duct.
We use capital letters to denote flow properties averaged in the
homogeneous spatial directions and in time, angle brackets to denote the averaging operator,
and lower-case letters to denote fluctuations from the mean.
We also exploit geometrical symmetries, and average the flow statistics over the duct quadrant or duct octant, whenever possible.
The effect of quadrant averaging was discussed in detail in our previous publication~\citet{pirozzoli_18}.

The $+$ superscript is here used to denote local wall units, namely quantities made nondimensional
with respect to the local friction velocity, $u_{\tau} = (\tau_w/\rho)^{1/2}$
(where $\tau_w= \rho \nu {\diff U}/{\diff y}\vert_w$ is the local wall shear stress),
and the local viscous length scale, $\delta_v=\nu/u_{\tau}$. 
The $*$ superscript is reserved to denote global wall units, with friction velocity 
based on the perimeter-averaged wall shear stress,
\begin{equation}
u^*_{\tau} = \left( \frac {\tau^*_w}{\rho} \right)^{1/2} = \left( \frac { h \left< \Pi \right> }{2 \rho} \right)^{1/2}, \label{eq:utaustar}
\end{equation}
and viscous length scale $\delta_v^*=\nu/u_{\tau}^*$.
Likewise, for normalization of the temperature field we consider either the local friction temperature,
\begin{equation}
\theta_{\tau} = \frac {q_w}{u_{\tau}}, 
\end{equation}
where $q_w = \alpha {\diff {\Theta}}/{\diff y}\vert_w$ is the local wall heat flux,
or the global friction temperature, 
\begin{equation}
\theta^*_{\tau} = \frac {q^*_w}{u^*_{\tau}} = \frac{ \left< {Q} \right> h}{2 u^*_{\tau}}.  \label{eq:ttaustar}
\end{equation}
Equations \eqref{eq:utaustar} and \eqref{eq:ttaustar} make it clear that the global friction velocity and temperature are related with the imposed forcing of the streamwise momentum and temperature equations,
rather than with the detailed distribution of wall momentum and temperature fluxes.
Notably, local and global wall units collapse in the case of canonical wall-bounded flows 
because of homogeneity, but the two are distinctly different in internal flows in non-circular 
ducts. DNS studies as the present one can then be used to distinguish between the two scalings,
which is impossible in more canonical set-ups.

\section{Temperature statistics}

\begin{figure}
 \begin{center}
(a)~\includegraphics[scale=0.12]{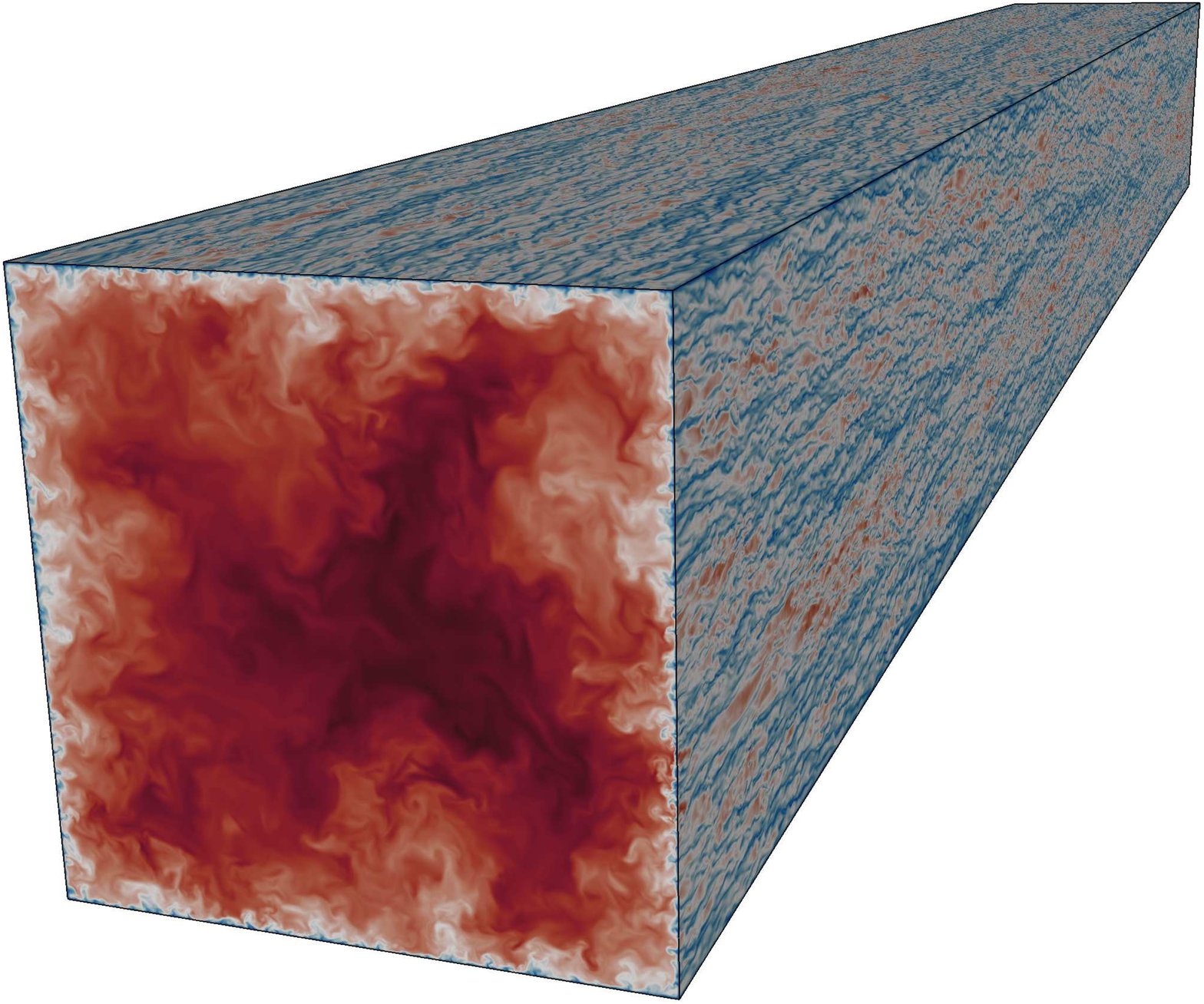}~(b)~\includegraphics[scale=0.12]{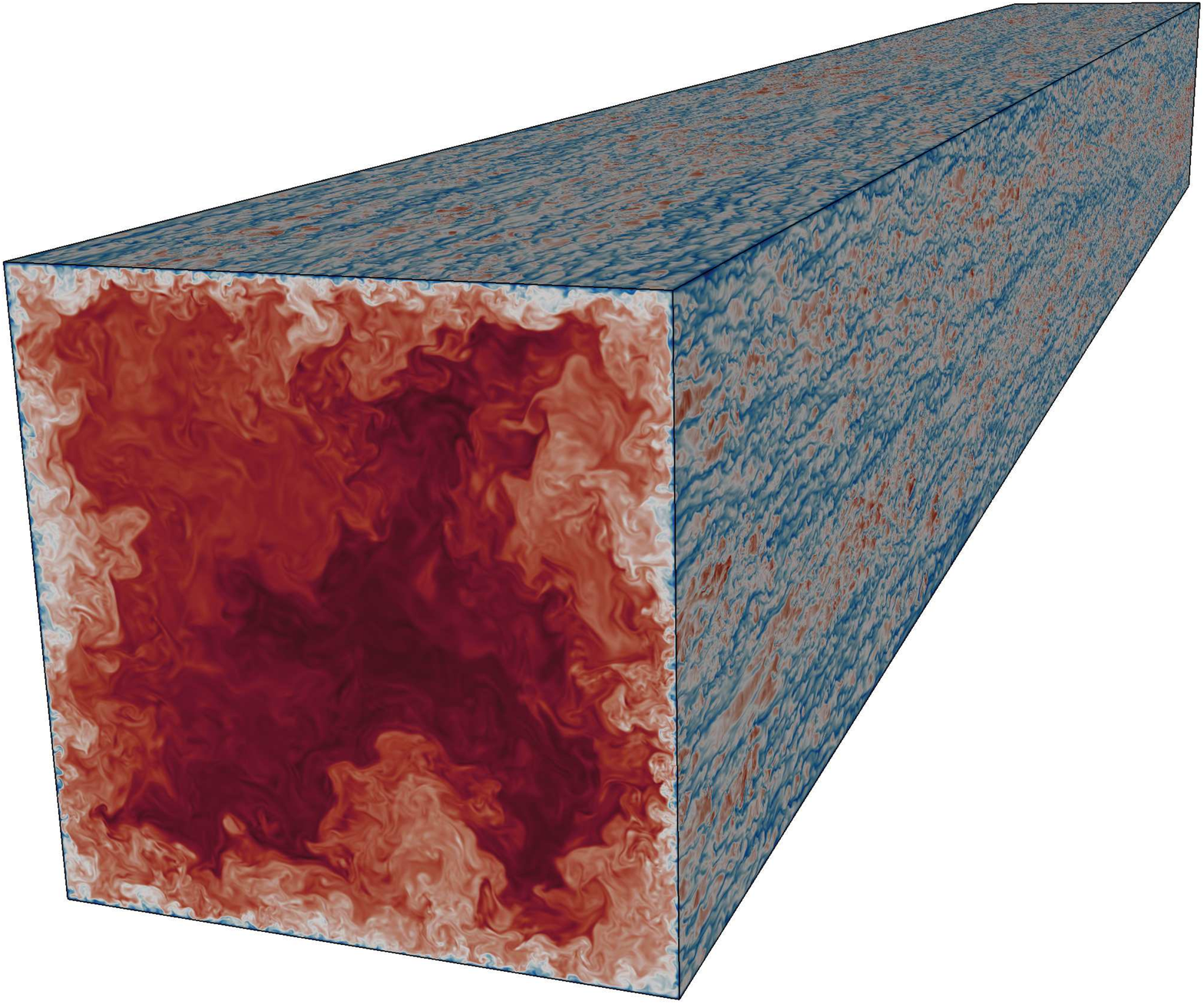}
  \caption{Instantaneous streamwise velocity (\textit{a}) and temperature (\textit{b}) fields for flow case F. Wall-parallel planes are taken at fifteen wall units from the walls.}
  \label{fig:phinst}
 \end{center}
\end{figure}

We begin by inspecting the instantaneous streamwise velocity and temperature fields in figure~\ref{fig:phinst}, 
which shows the flow both in the cross-stream and wall-parallel planes. 
The instantaneous flow structures are not different from those typical of canonical wall-bounded flows, and the near-wall region 
is populated with temperature and streamwise velocity streaks, resulting from sweep and ejections events visible in the cross-stream plane.
Large structures are also visible in the cross-stream plane which convey
high-speed, hot flow from the duct core towards the corners, and which interact with the smaller 
near-wall flow structures. Return motions of low-speed, cool fluid are also observed to emerge
from around the bisector of each side of the square duct. These observations
suggest that the secondary motions resulting from space- and time-averaging are not just an artifact,
but they are well present in the instantaneous flow realizations.
Similarity between streamwise velocity and temperature is apparent, with the main difference that
the latter reveals finer structures and sharper fronts owing to absence of the smoothing action of pressure~\citep{pirozzoli_16}.

\begin{figure}
 \begin{center}
  \includegraphics[scale=1.0]{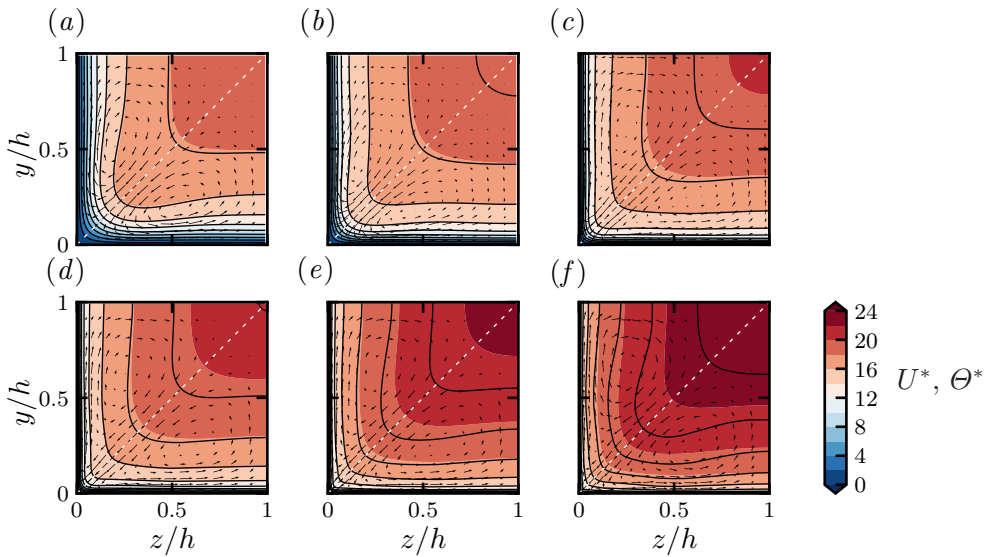}
	 \caption{Mean streamwise temperature (${\Theta}^*$, flooded contours), mean streamwise velocity (${U}^*$, lines), and  mean cross-stream velocity vectors ($V^*, W^*$) for flow cases A (\textit{a}), B (\textit{b}), C (\textit{c}), D (\textit{d}), E (\textit{e}), F (\textit{f}). For clarity, only a subset of the velocity vectors are shown. Only a quarter of the full domain is shown. The dashed diagonal lines indicate the corner bisector.}
  \label{fig:phimean}
 \end{center}
\end{figure}

\begin{figure}
 \begin{center}
  \includegraphics[scale=1.0]{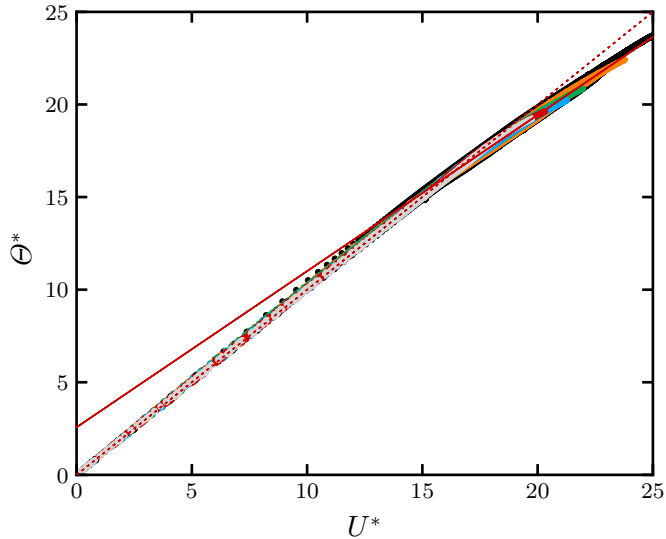}
  \caption{Scatter plot of mean temperature ($\Theta^*$) versus mean velocity ($U^*$) for DNS data of all flow cases, and corresponding 
   fitting functions for the near-wall region, $\Theta^* = U^*$ (dashed line), and for the outer region 
	 $\Theta^* = 2.57 + 0.84 \, U^*$ (solid line). Colors as in table~\ref{tab:test}.}
  \label{fig:scatter}
 \end{center}
\end{figure}

Figure~\ref{fig:phimean} depicts the mean temperature (contours) 
and velocity (iso-lines) fields in the duct cross section,
along with representative cross-stream velocity vectors. This representation 
brings to light the presence of a pair of counter-rotating secondary eddies in
each quarter of the domain, whose apparent role is bringing high-speed,
and high-temperature fluid from the duct core towards the corners, 
to compensate for the deficit.
As a result, both the temperature contours and the velocity isolines 
bend towards the corners, featuring a bulging first noticed by \citet{nikuradse_30}.
This effect seems to be non-monotonic with the Reynolds number, being 
most evident for DNS-A and DNS-E/F. 
As expected, temperature and velocity contours bear close
similarity, being nearly coincident near the wall, and retaining the same shape 
farther off. This analogy is the result of the formal similarity of the
controlling equations, and it is further investigated in figure~\ref{fig:scatter},
where we show a scatter plot of mean velocity and temperature. 
In the near-wall region (say, $U^* \le 10$), perfect matching is recovered
(namely, $\Theta^* = U^*$), which is not surprising, as at unit Prandtl number
$\Theta^+ \equiv U^+ \equiv y^+$ in the wall vicinity. 
However, near equality of the two distributions in global wall units is less trivial
and is a symptom of strong similarity between the two fields.
At non-unit Prandtl number we still expect a linear relationship to be present like
$\Theta^* = \Pran \, U^*$, since in general $\Theta^+ \equiv U^+ \equiv \Pran \, y^+$ in the wall vicinity~\citep{kader_81}.
Farther from walls ($U^* \gtrsim 15$, which corresponds to the root of the
logarithmic layer in canonical wall-bounded flows), a linear imprinting 
is still visible, although with larger scatter than in the near-wall region,
and with slope shallower than unity. This can be explained by recalling~\citep{pirozzoli_16}
that in canonical flows the distributions of both velocity and temperature are nearly logarithmic,
however with different slope. In particular, currently accepted values of the K\'arm\'an constant
are $\kappa \approx 0.39$ for the mean velocity, and $\kappa_{\theta} \approx 0.46$ for the mean temperature.
As a matter of fact, the scatter plot is well fitted with a linear function with slope
$\kappa/\kappa_{\theta} \approx 0.84$. At $\Pran \ne 1$ we thus expect a similar relationship,
however with different additive constant, as a result of the different intercept in the log law for the temperature field.
This observation is quite interesting in our opinion, as the presence of a 
(nearly) universal relationship between mean velocity and temperature in a non-trivial 
flow as a square duct would allow in principle to directly reconstruct the whole 
temperature field based on the velocity field.

\begin{figure}
 \begin{center}
  \includegraphics[scale=1]{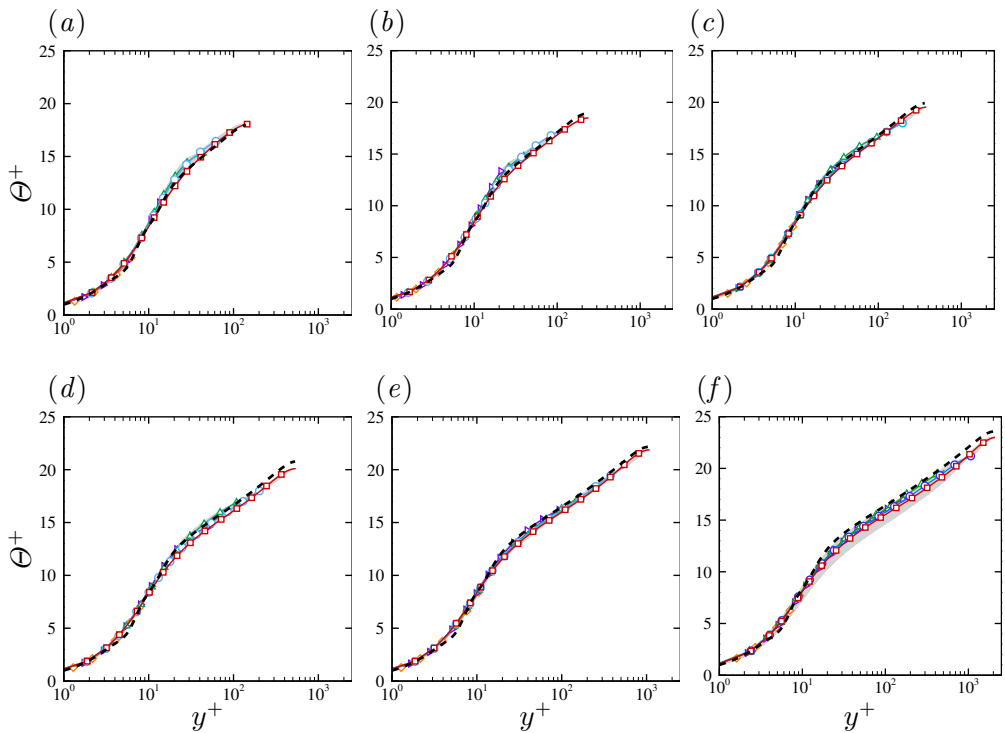}
  \vskip 1em
  \caption{Mean wall-normal temperature profiles scaled in local wall units for flow cases A (\textit{a}), B (\textit{b}), C (\textit{c}), D (\textit{d}), E (\textit{e}), F (\textit{f}). Profiles are plotted at several distances from the left wall, up to the corner bisector (see figure~\ref{fig:phimean} for reference): $z^*=15$ (diamonds), $z/h=0.1$ (right triangles), $z/h=0.25$ (triangles), $z/h=0.5$ (circles), $z/h=1$ (squares). The dashed lines denote fit of experimental data~\citep{kader_81}, at matching $\Rey_\tau^*$.}
  \label{fig:phi_prof}
 \end{center}
\end{figure}

\begin{figure}
 \begin{center}
  \includegraphics[scale=1]{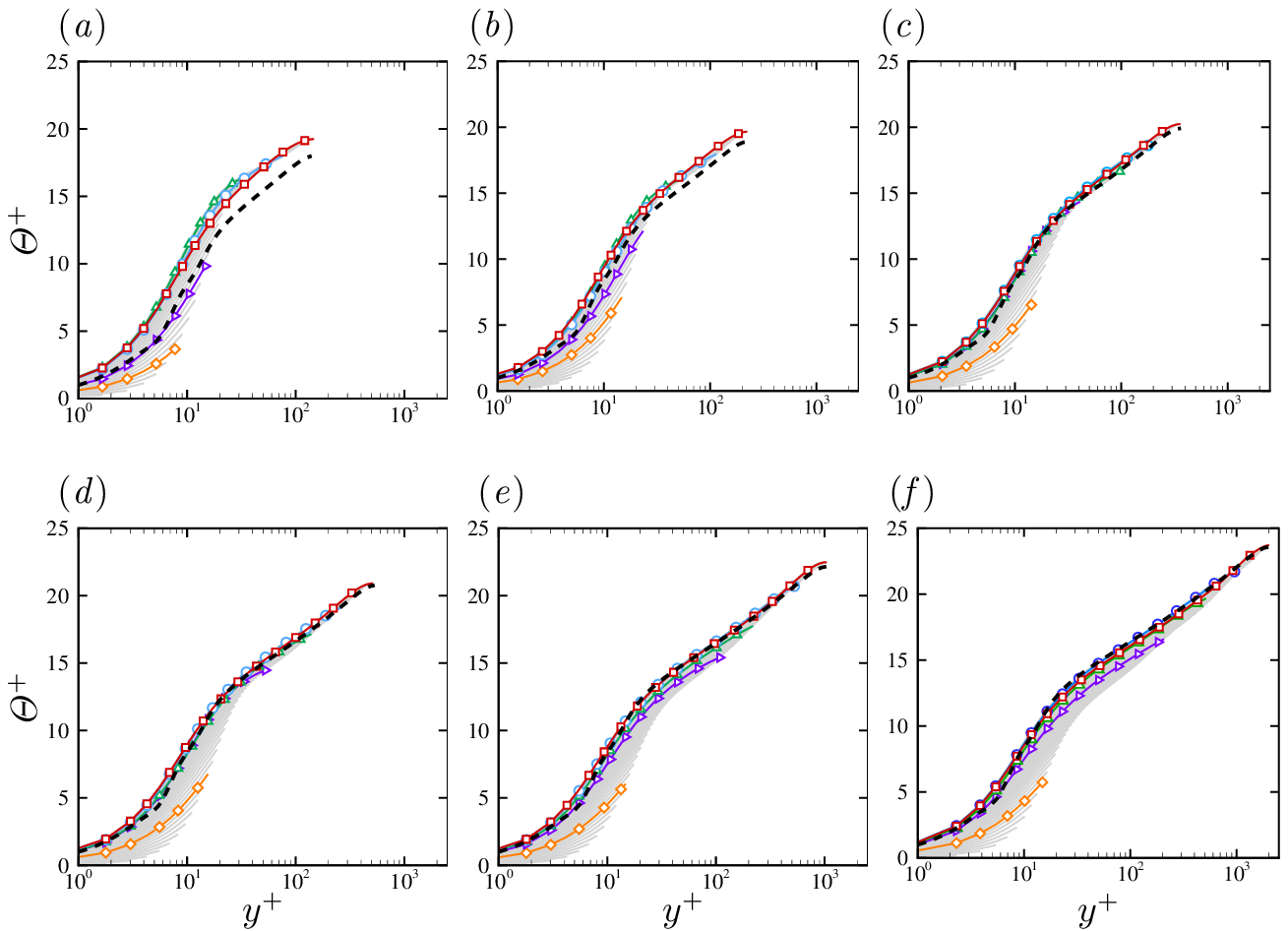}
  \vskip 1em
  \caption{Mean wall-normal temperature profiles scaled in global wall units for flow cases A (\textit{a}), B (\textit{b}), C (\textit{c}), D (\textit{d}), E (\textit{e}), F (\textit{f}). Profiles are plotted at several distances from the left wall, up to the corner bisector (see figure~\ref{fig:phimean} for reference): $z^*=15$ (diamonds), $z/h=0.1$ (right triangles), $z/h=0.25$ (triangles), $z/h=0.5$ (circles), $z/h=1$ (squares). The dashed lines denote fit of experimental data~\citep{kader_81}, at matching $\Rey_\tau^*$.}
  \label{fig:phi_prof_star}
 \end{center}
\end{figure}

Based on the previous observations, it is natural to study and compare the temperature statistics
expressed in local wall units ($+$) and in global wall units ($*$).
Figure~\ref{fig:phi_prof} shows the mean temperature profiles as a function of the
wall-normal distance up to the corner bisector (white dashed line in figure~\ref{fig:phimean}),
in local wall units. For reference purposes, the mean temperature profiles obtained the experimental
fitting by~\citet{kader_81} at matching $\Rey_{\tau}$ are also reported.
Excellent collapse of the locally scaled profiles is recovered near the wall,
also including the near-corner region.
The distributions become more widespread past $y^+ \approx 10$,
with maximum scatter of about $\pm 5\%$ in the ``logarithmic'' layer.
When the global friction temperature is used for normalization
(see figure~\ref{fig:phi_prof_star}),
scatter among the temperature profiles is observed
near the wall as a result of the local variation of the wall heat flux.
Perhaps unexpectedly, this normalization does yield similarly good universality
away from walls, and near coincidence with the pipe temperature profiles, 
at least at high enough Reynolds number.
This finding is probably related to the fact that mean temperature transport
away from solid walls is controlled by the imposed spatially uniform heat source
rather than by the nonuniform wall heat flux. 
Inspection of figure~\ref{fig:phi_prof_star} shows that transition from wall scaling
to 'global' scaling (which is controlled by the spatially uniform heat source)
occurs at a wall distance of about $0.2 h$,
which is also the lower limit for the core region in canonical flows~\citep{pope_00}.
This observation seems to support validity of Townsend's outer layer similarity also for the temperature field,
in that the outer flow mainly perceives the global heat flux and not its detailed local distribution at the wall.
Validity of outer-layer similarity for the temperature field was reported for rough walls~\citep{macdonald_19},
whereas we are not aware of similar conclusions for the case of more complex geometries because
most studies on forced convection are limited to canonical wall-bounded flows,
in which the wall heat flux is spatially uniform.
These results support findings previously reported for the streamwise velocity~\citep{pirozzoli_18},
and confirm that square duct flow is a convenient testbed for evaluating differences between local and global scaling.

\begin{figure}
 \begin{center}
  \includegraphics[scale=1.]{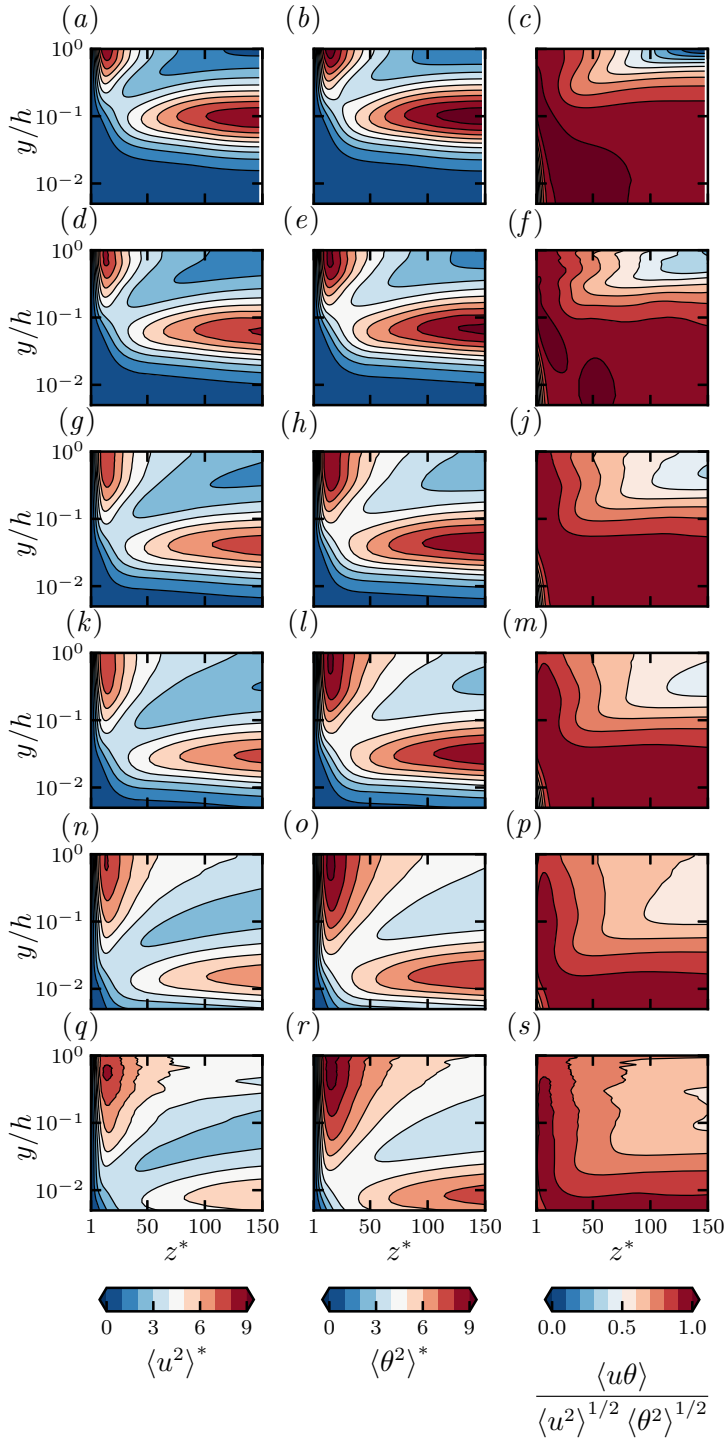}
  \vskip 1em
  \caption{Streamwise velocity and temperature fluctuation intensities (in global inner units), and their correlation coefficient,
   in the proximity of a duct corner for flow cases A (\textit{a}), (\textit{b}), (\textit{c});
   B (\textit{d}), (\textit{e}), (\textit{f}); C (\textit{g}), (\textit{h}), (\textit{j});
   D (\textit{k}), (\textit{l}), (\textit{m}); E (\textit{n}), (\textit{o}), (\textit{p});
   F (\textit{q}), (\textit{r}), (\textit{a}). Note that the horizontal coordinate ($z$) is given in inner units, and the vertical coordinate $y$ is given in outer units and 
   in a logarithmic scale.}
  \label{fig:urms}
 \end{center}
\end{figure}

The fluctuating velocity and temperature fields in the vicinity of 
the duct corners are shown in figure~\ref{fig:urms} upon global inner normalizations,
as well as the corresponding correlation coefficients.
To facilitate comparison across the Reynolds number range, 
the horizontal coordinate is reported in inner units up to $z^*=150$, and the vertical one in outer units. This allows simultaneous visualization
of both the corner and the core region, at all Reynolds numbers.
Far from the corner the distributions of the velocity and temperature variances 
are nearly unaffected by Reynolds number variation, and feature a buffer-layer peak at a distance of 
about fifteen wall units from the closer wall. 
The amplitude of the peak is reduced moving towards the corner,
whose effect is felt to within a distance of O(100) global wall units.
The $u-\theta$ correlation coefficient shows strict similarity of 
velocity and temperature fluctuations also in instantaneous sense, with 
the exception of the close neighborhood of the corner. 
Correlation in the outer layer is less than near the wall, but it tends to increase 
at higher Reynolds number, similar to the case of plane channel flow~\citep{pirozzoli_16}.

\begin{figure}
 \begin{center}
  \includegraphics[scale=1.0]{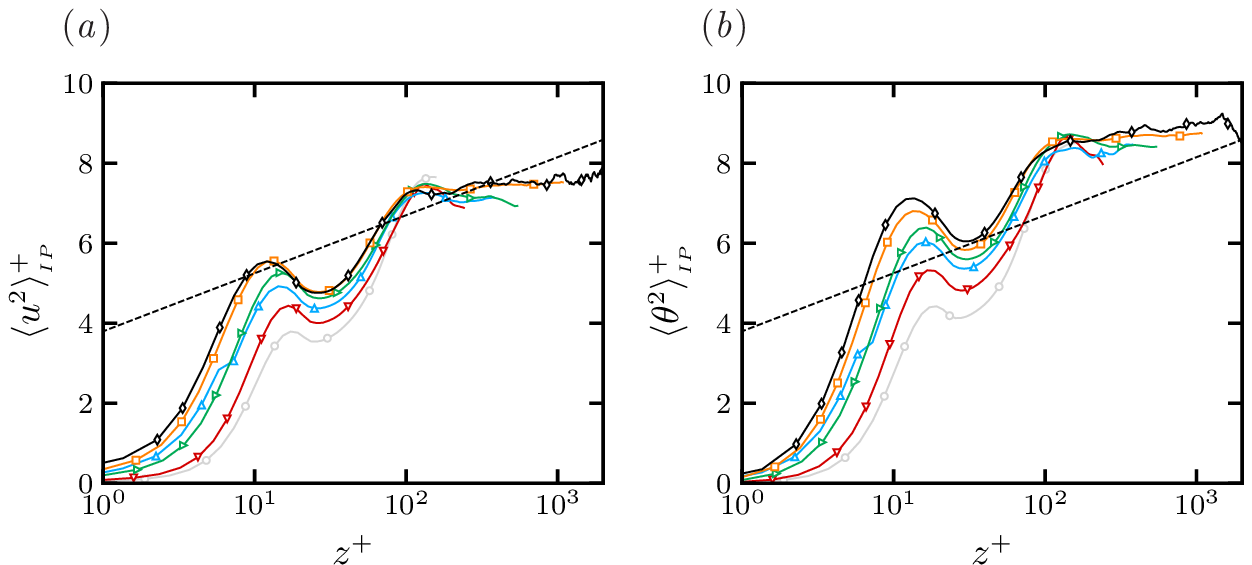}
	 \caption{Inner-layer peak variances of axial velocity (\textit{a}), and temperature (\textit{b}),
  expressed as a function of the local friction Reynolds number ($z^+$).
  The dashed lines denote the logarithmic fit for canonical flows~\citep{marusic_17}, namely
	 $\langle u^2\rangle^+_{_{IP}} = 0.63 \log \Rey_{\tau} + 3.8$. Symbols as in table~\ref{tab:test}.}
  \label{fig:upk}
 \end{center}
\end{figure}

A quantitative analysis of the buffer-layer peaks of the velocity and temperature 
variances is reported in figure~\ref{fig:upk}. Specifically, for each horizontal location 
($z$, measured from the corner), 
we show the peak variances of $u$ and $\theta$ expressed in local wall units.
In this respect we note that $z^+$ can be interpreted as 
effective local friction Reynolds numbers, based on the distance of the 
wall point from the corner bisector (see figure~\ref{fig:phimean}). 
It is known that in canonical flows~\citep{marusic_13}, the peak velocity variances
increase logarithmically with the friction Reynolds number. Hence, it is interesting 
to verify whether the same trend also occurs in flows involving non-trivial cross sections,
and featuring secondary motions.
For that purpose, in all panels of figure~\ref{fig:upk} we show the logarithmic 
trend predicted by \citet{marusic_13} as benchmark. 
It is noteworthy that, away from the corner (say $z^+ \gtrsim 100$) local wall units yield
good universality of the distributions, and suggest an increasing trend 
of the variances with the local friction Reynolds number. 
However, the growth rate seems to be far less than in canonical
flows, possibly violating of Townsend's attached-eddy hypothesis.
We speculate that this peculiar behavior may be either due to 
disruption of the hierarchy of wall-attached eddies due to the action of secondary motions,
or perhaps to impeded spanwise growth caused by the presence of 
corners, both effects being absent in planar channel or pipe flow.

\begin{figure}
 \begin{center}
  \includegraphics[scale=1.,clip]{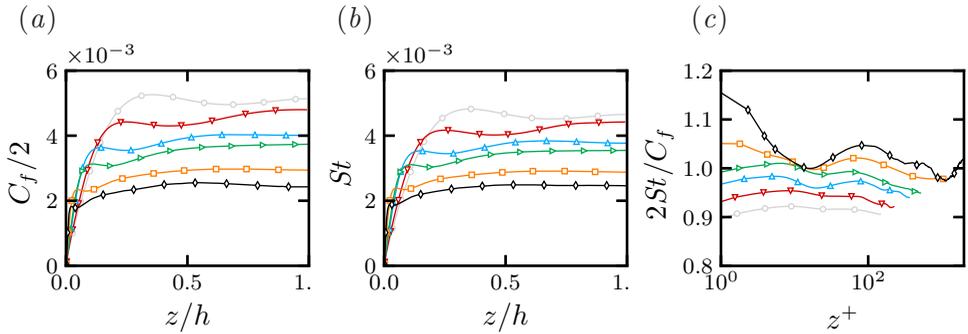}
	 \caption{Distributions of local wall friction coefficient $C_f=2 u_{\tau}^2/u_b^2$ (\textit{a}), local Stanton number $\St=(u_{\tau} \theta_{\tau})/(u_b \theta_m)$ (\textit{b}), and Reynolds analogy factor (\textit{c}) along left part of lower wall.
   Symbols as in table~\ref{tab:test}.}
  \label{fig:wallq}
 \end{center}
\end{figure}

The distributions of the local friction and heat transfer coefficients along
the duct wall are shown in figure~\ref{fig:wallq}. As previously shown by \citet{pirozzoli_18},
the distributions of the wall friction coefficients is non-monotone,
featuring a peak at the side bisector, and a secondary peak 
near the corner, at a distance scaling with global wall units.
As from theoretical predictions~\citep{spalart_18},
the friction tends to equalize over the duct perimeter 
at higher $\Rey$. The heat transfer coefficient seems 
to have the same qualitative behavior, thus corroborating 
the experimental observations of \citet{brundrett_67, hirota_97}.
Quantitative assessment is shown in figure~\ref{fig:wallq}(c), where we show the 
Reynolds analogy factor, $2 \St / C_f$, which is expected 
to be unity at $\Pran = 1$~\citep{white_06}. 
In fact, whereas deviations are observed at low Reynolds number, 
the distribution corresponding to DNS-F varies between 0.98 and 1.2,
right at the corner.
 
\begin{figure}
 \begin{center}
  \includegraphics[scale=1,clip]{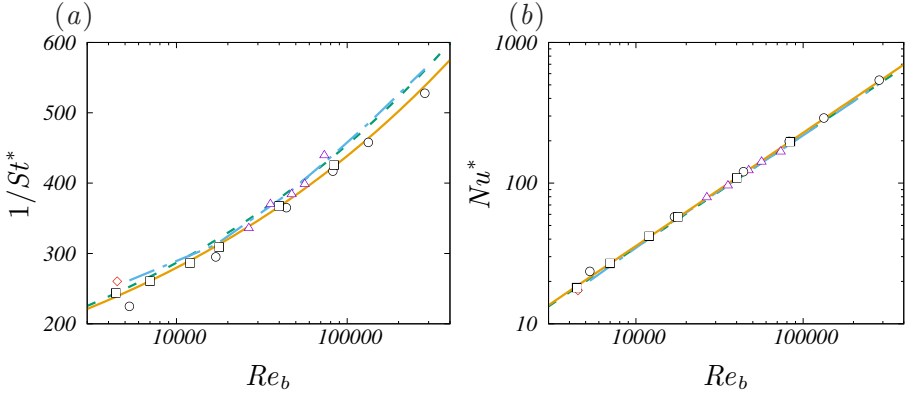}
  \vskip 1em
  \caption{
	  Inverse global Stanton number (\textit{a}) and Nusselt number (\textit{b}) as a function of bulk Reynolds number,
           from the present DNS (squares), from DNS of circular pipe flow~\citep[circles][]{pirozzoli_21},
           from previous LES~\citep[diamonds][]{pallares_02}, 
           and from experiments~\citep[][triangles]{brundrett_67}.
           The dashed line denotes the correlation \eqref{eq:KC},
           and the dot-dashed lines the correlation \eqref{eq:gnielinski}.
           }
  \label{fig:Nu}
 \end{center}
\end{figure}

The overall heat transfer performance of the duct is quantified in terms of the global Stanton number,
\begin{equation} 
\St^* = \frac{q_w^*}{u_b \theta_m}
= \frac 1{u_b^* \theta_m^*}, \label{eq:stanton}
\end{equation}
or more frequently in terms of the Nusselt number,
\begin{equation}
\Nu^* = \St^* \, \Rey_b \, \Pran . \label{eq:nusselt}
\end{equation}
In figure~\ref{fig:Nu} we compare the distributions of the heat transfer coefficients
(inverse Stanton number and Nusselt number) obtained from the present DNS (squares) with those resulting from DNS of circular pipe flow (circles), 
from previous LES (diamonds), and from experiments (triangles) in square ducts.
We note that the latter two datasets were obtained for $\Pran=0.71$, hence
the $\Nu$ data are rescaled by a factor $1/\sqrt{\Pran}$ to compare with the present ones.
As a reference, we also report correlations widely used in the engineering practice,
including that by \citet{gnielinski_75},
\begin{equation}
\Nu = \frac {\Pran C_f / 32 \left( \Rey_b - 1000 \right)}{1 + 12.7 (C_f/32)^{1/2} \left( \Pran^{2/3} - 1\right)}, \label{eq:gnielinski}
\end{equation}
and from \citet{kays_93},
\begin{equation}
\Nu = 0.022 \, \Rey_b^{0.8} \, \Pran^{0.5}. \label{eq:KC} 
\end{equation}
Last, direct fitting the present DNS data (at $\Pran=1$) with a power-law expression yields
\begin{equation}
\Nu = 0.0216 \, \Rey_b^{0.805}. \label{eq:powerlaw} 
\end{equation}

The DNS data show good correspondence with the pipe DNS data 
at matching $\Rey_b$, with the exception of the lower Reynolds number case,
which supports validity of the hydraulic diameter concept as the
relevant length scale for heat transfer data reduction.
Agreement with experiments and LES in square ducts is also quite good,
on account of experimental uncertainties and modeling errors in LES.
It is interesting that differences are levelled off when the popular representation
in terms of the Nusselt number is used, as in figure~\ref{fig:Nu}(b), hence we 
believe that the $1/\St$ representation should be used when relatively small differences
must be discriminated.
Classical correlations seem to suggest systematic difference of up to $5\%$ 
in the prediction of the heat transfer coefficient. 
This difference may be partly due to inaccuracy of correlations based on old experimental data,
or to the fact that those are mainly trained for the $\Pran=0.71$ case, whereas here $\Pran=1$.
Discrepancies can also be partly attributed to our heat forcing scheme, in which a spatially uniform
heat source is prescribed, which tends to slightly overpredict
the heat flux as compared to other approached in which the 
total heat flux is maintained strictly constant in time~\citep{abe_17,alcantara_21}.
Slight adjustment of the Kays--Crawford power-law formula coefficients 
as after equation~\eqref{eq:powerlaw} seems to yield very good representation of the DNS data.

An issue which deserves further investigation is why use of the hydraulic diameter
yields excellent results, at least in the case of square ducts under consideration.
\citet{pirozzoli_18} showed that universality of friction is connected with 
near applicability of the logarithmic velocity law in the direction normal to the
nearest wall. It is worthwhile verifying whether it is also the case of the heat flux coefficient.
As previously shown, the temperature distributions 
are nearly universal away from walls, even when expressed in
global wall units. 
Hence, approximating the outer layer profiles with the classical log law,
namely $U^*=1/{\kappa} \log y^* + C$, $\Theta^*=1/\kappa_{\theta} \log y^* + C_{\theta}$, 
and integrating over the duct cross section,
the following expression for the inverse Stanton number velocity results
\begin{equation}
\begin{split}
1 / \St^* = u_b^* \theta_m^* = \frac{8}{4\Rey_\tau^{*2}}\int_0^h\int_0^z\left[\frac{1}{\kappa \kappa_\theta}\log{y^*}^2 + \log{y^*}\left(\frac{C_\theta}{\kappa} + \frac{C}{\kappa_\theta}\right) +CC_\theta\right]dy^*dz^* = \\
\frac{1}{\kappa \kappa_\theta}\left(\log^2{\Rey_\tau^*} + 3\log{\Rey_\tau^*} + \frac{7}{2}\right) + \left(\frac{C_\theta}{\kappa}+\frac{C}{\kappa_\theta}\right)\left(\log{\Rey_\tau^*}-\frac{3}{2}\right) + CC_\theta .
\label{eq:stinv}
\end{split}
\end{equation}
Equation~\eqref{eq:stinv} should be compared with the corresponding expression for
a circular duct with diameter $D$,
\begin{equation}
1 / \St = \frac{1}{\kappa \kappa_\theta}\left(\log^2{\Rey_\tau} + 3\log{\Rey_\tau} + \frac{7}{2}\right) + \left(\frac{C_\theta}{\kappa}+\frac{C}{\kappa_\theta}\right)\left(\log{\Rey_\tau}-\frac{3}{2}\right) + CC_\theta ,
\end{equation}
where $\Rey_{\tau} = D u_{\tau} / (2 \nu)$.
The two expressions are identical for $2 h = D$, hence provided
the Reynolds number based on the hydraulic diameter is the same.
It is interesting that equation~\eqref{eq:stinv} is basically arrived at
by neglecting the local wall shear stress and heat flux variation along the duct perimeter,
and disregarding the flow deceleration at corners.
Apparently, these effects very nearly cancel out upon integration.

\section{Contribution of secondary flows to heat transfer}

In order to quantify the effects of secondary motions on heat transfer
we derive a generalized version of the Fukagata--Iwamoto--Kasagi (FIK) identity~\citep{fukagata_02} for the
Stanton number, by following the approach of~\citet{modesti_18}.
We consider the mean temperature balance equation,
\begin{equation}
\alpha \nabla^2 \Theta = \nabla \cdot\boldsymbol{q}_C + \nabla\cdot\boldsymbol{q}_T
- \langle{Q}\rangle ,
\label{eq:mtb}
\end{equation}
where
$\boldsymbol{q}_C={\Theta}\,{\mathbf{U}}_{yz}$ is associated with mean cross-stream convection (hence, with the secondary motions),
$\boldsymbol{q}_T={\langle \theta \mathbf{u}_{yz} \rangle}$ is associated with turbulence convection,
$\mathbf{u}_{yz}=\left({v},{w}\right)$ is the cross-stream velocity vector,
and $\langle{Q}\rangle$ is the mean driving bulk heating.

Equation~\eqref{eq:mtb} may be interpreted as a Poisson equation for the mean temperature, with source terms
$\nabla\cdot\boldsymbol{q}_T$, $\nabla\cdot\boldsymbol{q}_C$ and $\langle Q\rangle$ obtained from the DNS dataset.
Hence, the solution of equation \eqref{eq:mtb} may be formally cast as the superposition of three parts,
namely ${\Theta}={\Theta}_D+{\Theta}_T+{\Theta}_C$, and these temperature fields can be obtained by solving three Poisson equations,
\begin{equation}
\alpha \nabla^2{\Theta}_D=-\langle{Q}\rangle, \quad \alpha \nabla^2{\Theta}_T=\nabla\cdot\boldsymbol{q}_T, \quad \alpha \nabla^2{\Theta}_C=\nabla\cdot\boldsymbol{q}_C,
\label{eq:fik_temp}
\end{equation}
with homogeneous boundary conditions,
where ${\Theta}_D$, ${\Theta}_T$, and ${\Theta}_C$ denote the diffusive, turbulent, and convective contributions to the mean temperature field.
The mean temperature in the duct may accordingly be evaluated as
\begin{equation}
\theta_b=\frac {\langle{Q}\rangle A}{\alpha} {\theta_b}_1+{\theta_b}_T+{\theta_b}_C, \quad {\theta_b}_X = 
\frac{1}{A} \int_A {\Theta}_X\mathrm{d}A, \label{eq:tmean}
\end{equation}
where we have introduced the unitary Stokes temperature field $\theta_1$, defined as solution of $\nabla^2 \theta_1 = -1/A$, which
by construction is only a function of the duct geometry, and which allows us to express the diffusion-associated temperature
field as $\Theta_D= A \, \theta_1 \langle{Q}\rangle / \alpha$.
For convenience, in the present analysis we use a modified form of the Stanton number,
based on the mean temperature, rather than on the bulk temperature, namely 
\begin{equation}
\St_b^* = \frac{q_w^*}{u_b \theta_b}
= \frac 1{u_b^* \theta_b^*}, \label{eq:stanton_b}
\end{equation}
instead of \eqref{eq:stanton}, which when substituted into equation~\eqref{eq:tmean} yields
\begin{equation}
\St_b^* = \frac{1}{\Pran \Rey_P {\theta_b}_1}\left(1-\frac{{\theta_b}_T}{\theta_b} - \frac{{\theta_b}_C}{\theta_b} \right)=\St_D + \St_T + \St_C\quad,
\label{eq:fik}
\end{equation}
where $\Rey_P=u_b P / \nu$ is the bulk Reynolds number based on the duct perimeter $P$.
Equation~\eqref{eq:fik} clearly shows that the heat transfer coefficient may be regarded as
the sum of contributions due to diffusivity, turbulence, and convection (labeled as ${St}_X$ in~\eqref{eq:fik}),
and it yields a similar expression to classical FIK identity for the friction coefficient.

\begin{table}
\footnotesize
\centering
\begin{tabular}{cccccccc}
\hline
 Case   &   $\St^*_V \times 10^3$ &   $\St^*_T \times 10^3$ &   $\St^*_C \times 10^3$ &   $\St^* \times 10^3$ &   $\St^*_V/\St^*(\%)$ &   $\St^*_T/\St^*(\%)$ &   $\St^*_C/\St^*(\%)$ \\
\hline
 A      &                  0.095 &                  0.144 &                  0.030 &                0.268 &                35.3 &                53.7 &                11.0 \\
 B      &                  0.060 &                  0.176 &                  0.009 &                0.245 &                24.4 &                72.0 &                 3.6 \\
 C      &                  0.035 &                  0.172 &                  0.009 &                0.216 &                16.2 &                79.8 &                 4.0 \\
 D      &                  0.024 &                  0.170 &                  0.007 &                0.201 &                11.7 &                84.7 &                 3.6 \\
 E      &                  0.010 &                  0.149 &                  0.008 &                0.167 &                 6.2 &                89.2 &                 4.5 \\
 F      &                  0.005 &                  0.123 &                  0.010 &                0.138 &                 3.6 &                89.4 &                 7.0 \\
\hline
\end{tabular}
\caption{
Contributions of diffusion, turbulence and mean convection terms to the heat transfer coefficient, as from equation~\eqref{eq:fik}.
}
\label{tab:fik}
\end{table}

\begin{figure}
 \begin{center}
  \includegraphics[scale=1]{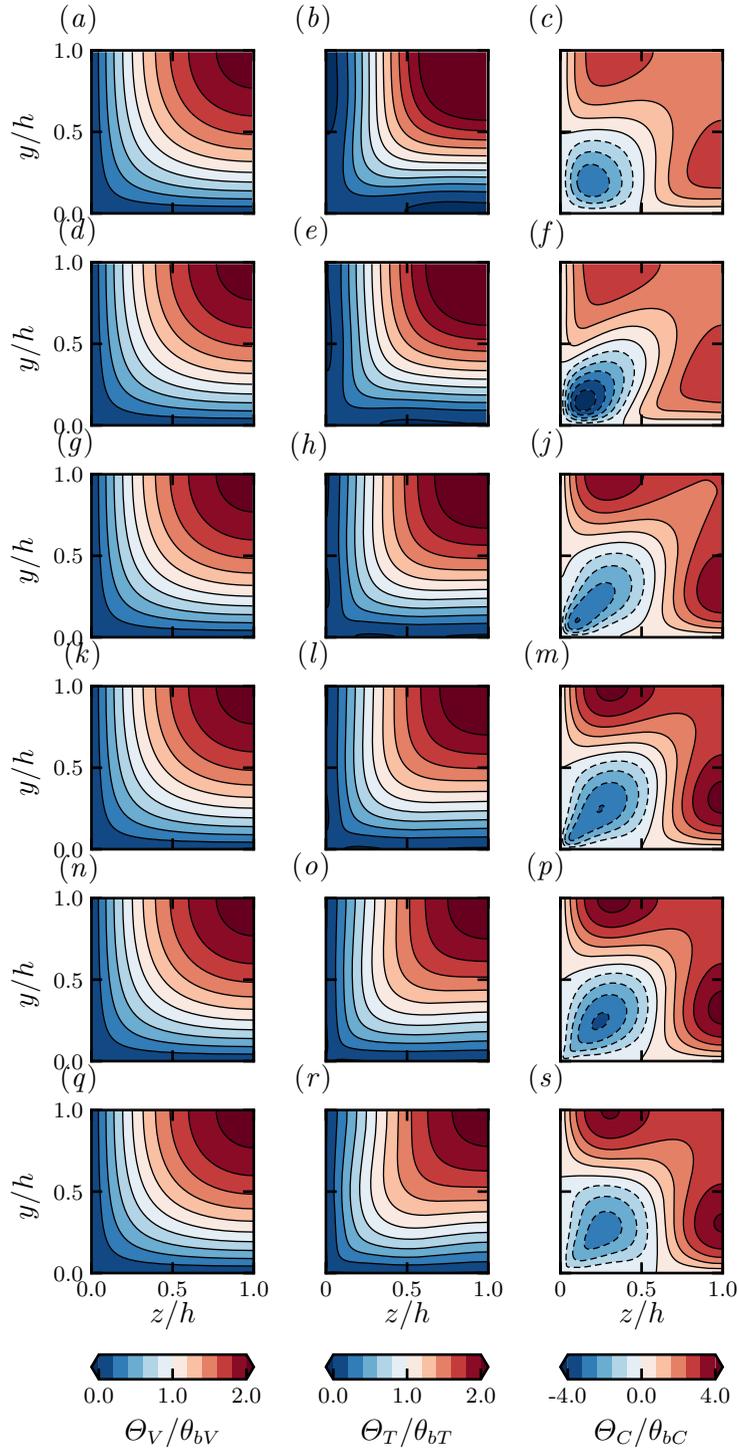}
  \caption{Contribution of diffusion, turbulence and mean convection terms to the mean temperature field, as determined by solving equation~\eqref{eq:fik_temp}, for flow cases A (\textit{a}), (\textit{b}), (\textit{c});
   B (\textit{d}), (\textit{e}), (\textit{f}); C (\textit{g}), (\textit{h}), (\textit{j});
   D (\textit{k}), (\textit{l}), (\textit{m}); E (\textit{n}), (\textit{o}), (\textit{p});
   F (\textit{q}), (\textit{r}), (\textit{a}).}
  \label{fig:phifik}
 \end{center}
\end{figure}

We note that this formalism has additional generality than the classical FIK identity
and its extensions~\citep{peet_09,jelly_14},
and it allows to isolate the effects of mean convection,
turbulent and diffusion terms to the wall shear stress distributions along the duct perimeter, resulting from
\begin{equation}
{q_w}_D = \alpha \left.\frac{\partial {\Theta}_D}{\partial y}\right\rvert_w, \quad 
{q_w}_T = \alpha \left.\frac{\partial {\Theta}_T}{\partial y}\right\rvert_w, \quad
{q_w}_C = \alpha \left.\frac{\partial {\Theta}_C}{\partial y}\right\rvert_w. \label{eq:qw_fik}
\end{equation}
Regarding equation~\eqref{eq:qw_fik} it is important to note that only
${q_w}_D$ has non-zero mean, whereas integration of equation~\eqref{eq:fik_temp} 
for ${\Theta}_C$ and ${\Theta}_T$ readily shows
that their integrated contributions vanish as $\boldsymbol{q}_C$, $\boldsymbol{q}_T$ are both zero at walls.
The contributions to the mean heat transfer coefficients are given in table~\ref{tab:fik},
both in absolute terms and as a fraction of the total.
Consistent with physical expectations,
the diffusion contribution is observed to decline at increasing Reynolds number with respect to
the turbulent term. The contribution of cross-stream convection is found to be roughly constant
across the explored Reynolds number range, however it remains much less than the turbulence
contribution. Extrapolating the DNS data, we expect the convective contribution to exceed the diffusion
one at high enough Reynolds number.
Figure~\ref{fig:phifik} shows the mean temperature fields associated with the diffusion, 
turbulent and convective terms. For the sake of clarity, each contribution 
is normalized with the corresponding bulk value, ${\theta_b}_X$,
where ${\theta_b}_T$ and ${\theta_b}_C$ are both negative, thus providing an additive effect to the
heat transfer (see equation~\eqref{eq:fik}).
The diffusion-associated temperature field $\Theta_V$ arises from the solution of a Poisson equation with uniform right-hand-side
(the heat source), hence its shape is identical to the case of laminar flow~\citep{shah_14},
only depending on the duct cross-sectional geometry.
The turbulence-associated temperature field $\Theta_T$, 
is everywhere negative and topologically similar to the diffusion-associated field, highlighting 
the cooling effect of turbulence on the bulk flow, with incurred increase of the heat transfer coefficient.
The temperature field $\Theta_C$ associated with mean convection 
has a more complex organization. Positive values are found near the duct corners, whereas negative values are found
near the duct bisectors, the zero crossings being located half-way in between. This finding is consistent with the intuitive expectation
that secondary motions tend to equalize temperature across the duct cross-section, thus quantitatively corroborating
claims made by early investigators~\citep{prandtl_27}.
Notably, all the distributions shown in figure~\ref{fig:phifik} are not qualitatively affected by Reynolds number variation when scaled with respect to their mean integral value, thus showing that change of the
Reynolds number only changes the relative importance of the three terms, as quantified in table~\ref{tab:fik}.

\begin{figure}
 \begin{center}
  \includegraphics[scale=1]{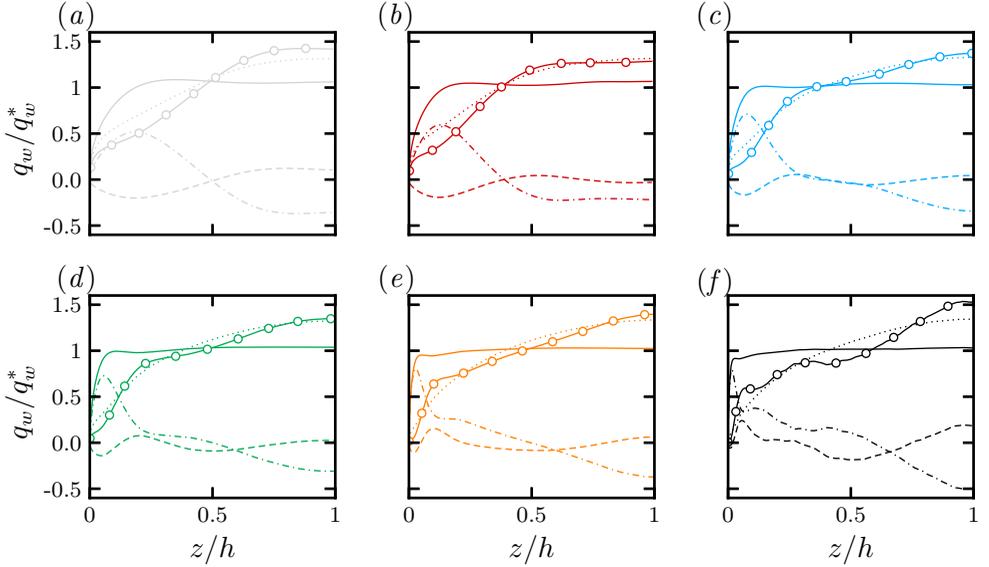}
  \caption{Contributions to mean wall heat flux (shown along half duct side)
	   for flow cases A (\textit{a}), B (\textit{b}), C (\textit{c}), D (\textit{d}), E (\textit{e}), F (\textit{f}):
           diffusive (${q_w}_D(z)$, dotted),
           turbulent (${q_w}_T(z)$, dashed),
           diffusive + turbulent (${q_w}_D(z)+{q_w}_T(z)$, circles),
           mean convection (${q_w}_C(z)$, dash-dotted),
           and total ($q_w(z)$, solid line).}
  \label{fig:qw_fik}
 \end{center}
\end{figure}

The observations made regarding the organization of the temperature field have direct impact on
the distribution of the local wall heat flux, shown in figure~\ref{fig:qw_fik}.
The heat flux distribution induced by diffusion terms is the same for all cases, and nearly parabolic in shape.
The turbulence terms have a complex behavior, exhibiting multiple peaks which
change with the Reynolds number. In general, they yield heat flux increase toward the duct bisectors,
and slight attenuation at the corners (compare the lines with circles and with dots), thus
making the heat flux distributions more nonuniform. Mean convection 
yields large positive peaks in the corner vicinity, whose distance
from the neighboring walls scales in inner units~\citep{pirozzoli_18}, and negative values
around the duct bisector. As a result, the distributions of the heat flux 
tend to be rather flat, especially at higher Reynolds number.

\begin{figure}
 \begin{center}
  \includegraphics[scale=0.9]{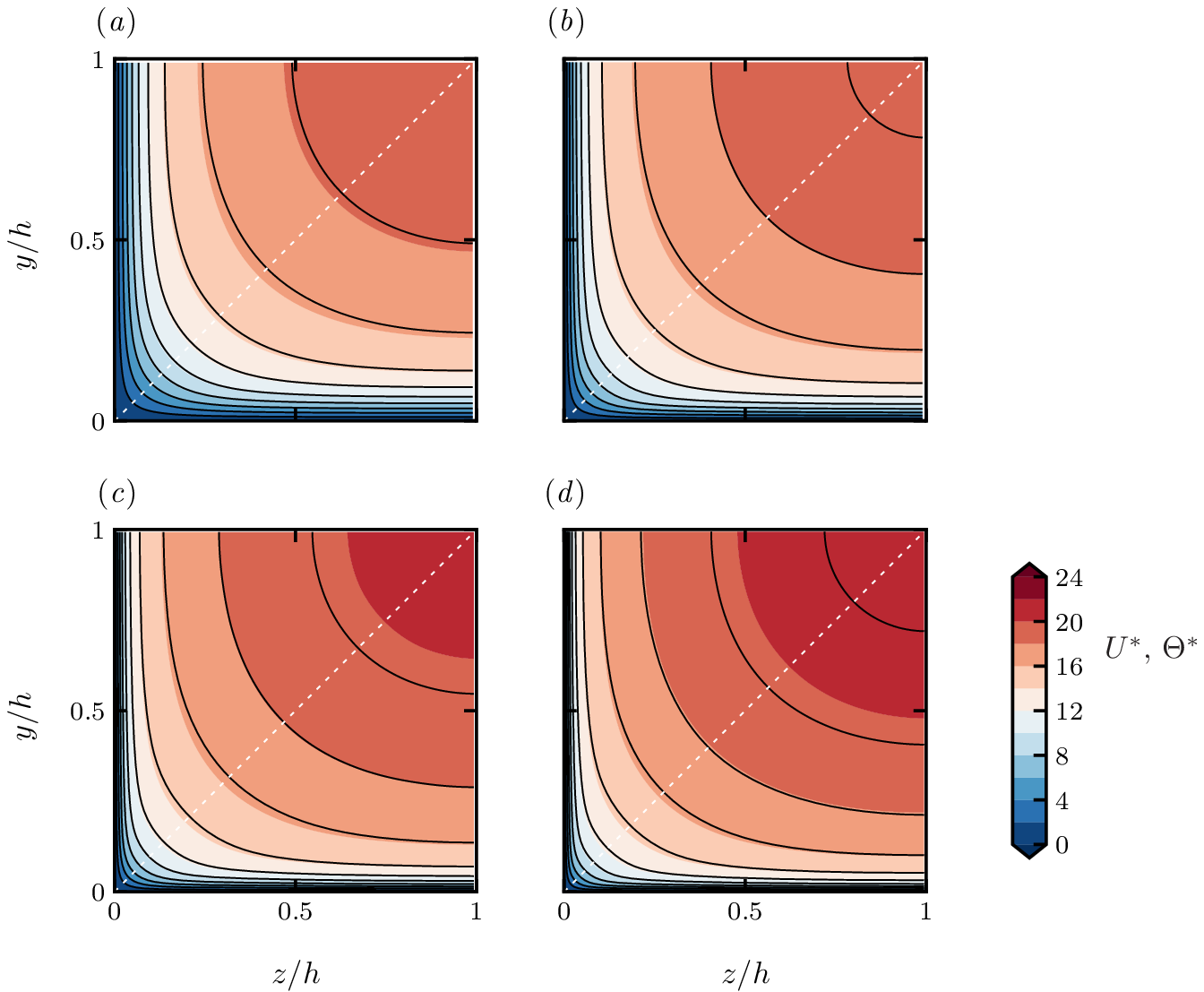}
	 \caption{Flow cases without secondary motions: mean streamwise velocity (${U}^*$, lines), and mean temperature (${\Theta}^*$, flooded contours), for flow cases A0 (\textit{a}), B0 (\textit{b}), C0 (\textit{c}), D0 (\textit{d}). 
	  Only a quarter of the full domain is shown. The dashed diagonal lines indicate the corner bisector.}
  \label{fig:phimean_novw}
 \end{center}
\end{figure}

As an alternative to quantify the effect of secondary flows we have also considered the approach
of~\citet{modesti_18}, and carried out numerical experiments
at the same bulk Reynolds numbers as the baseline cases, whereby the secondary motions
are artificially suppressed (flow cases with the 0 suffix in table~\ref{tab:test}). 
For that purpose we force the streamwise-averaged cross-stream velocity components to have zero mean by setting
\begin{subequations}
 \begin{align}
  v(x,y,z,t) &\rightarrow v(x,y,z,t) - \overline{v}^x(y,z,t)\\
  w(x,y,z,t) &\rightarrow w(x,y,z,t) - \overline{w}^x(y,z,t),
 \end{align}
\end{subequations}
at each Runge-Kutta sub-step,
where $\overline{(.)}^x$ denotes the streamwise averaging operator.
Figure~\ref{fig:phimean_novw} shows the temperature contours with superposed streamwise velocity isolines for flow cases A0--D0.
We note that the typical bulging of velocity and temperature towards the corners is absent for these cases,
and the core region shows near axial symmetry as in circular pipe flow.
The mean velocity and temperature fields are nearly coincident at low Reynolds number, whereas differences
become more evident as $\Rey_b$ increases. 
Although the mean flow fields in figure~\ref{fig:phimean_novw} 
obviously differ from the full DNS cases shown in figure~\ref{fig:phimean},
the heat transfer is found to be barely affected.
In fact, similar to what reported by~\citet{modesti_18} for frictional drag, suppressing the secondary flows
yields even higher hear transfer at low Reynolds number, whereas bare $3\%$ decrease of the Nusselt number 
is found for flow case $D0$, see table~\ref{tab:test}. 
These figures are consistent with those suggested by the FIK identity, 
which actually returns about $3\%$ contribution of secondary flows to
global heat transfer for flow case D.

\section{Modeling the turbulent heat flux}

The numerical solution of the Reynolds Averaged Navier--Stokes (RANS) equations is a common approach for flows of industrial interest.
Solution of the temperature field requires a suitable closure models for the turbulent heat flux vector,
$\boldsymbol{q}=\langle{\boldsymbol{u}\theta}\rangle$.
Such closures are typically less sophisticated than those developed for momentum flux,
and generally relying on validity of the Reynolds analogy.
The primary outcome of this school of thought is the popularity retained over the years by the turbulent Prandtl number,
\begin{equation}
 \Pran_t=\frac{\nu_t}{\alpha_t}, 
 \label{eq:prt}
\end{equation}
where $\nu_t$ and $\alpha_t$ are the eddy viscosity and eddy diffusivity coefficients, respectively.
These coefficients are the heart of the linear eddy-viscosity hypothesis and its heat-transfer counterpart,
\begin{equation}
\widetilde{\boldsymbol{a}} = -\nu_t \left( \nabla \boldsymbol{U} + \nabla \boldsymbol{U}^T \right), \quad 
\widetilde{\boldsymbol{q}} = -\alpha_t \nabla \Theta, \label{eq:linear_eddy}
\end{equation}
where $\widetilde{\boldsymbol{a}}$ is the modelled counterpart of the anisotropic Reynolds stress tensor ($\boldsymbol{a}=\boldsymbol{\tau}-2/3 k$), 
and $\widetilde{\boldsymbol{q}}$ the modelled turbulent heat flux vector, with
$\boldsymbol{\tau}=\langle{\boldsymbol u \boldsymbol u}\rangle$, $k=1/2 \, \mathrm{tr} (\boldsymbol{\tau})$.
Most heat transfer models assume uniform turbulent Prandtl number ($\Pran_t \approx 1$) to avoid solving ad-hoc transport equations
for $\alpha_t$, although this hypothesis is not always accurate~\citep{pirozzoli_16,kaller_19}.
Even assuming that the approximation $\Pran_t \approx 1$ holds,
limitations of the linear eddy-viscosity hypothesis~\eqref{eq:linear_eddy} are evident,
and reliance on the turbulent Prandtl number concept replicates modeling inaccuracies.
\citet{pope_75} extended the eddy-viscosity ansatz by using a tensor polynomial to model $\widetilde{\boldsymbol{a}}$. 
This approach is mathematically sound because it relies on the Cayley--Hamilton theorem.
In fact, it turns out that five tensor bases are in most cases sufficient to recover exact representation of 
$\widetilde{\boldsymbol{a}}$~\citep{gatski_93,jongen_98,modesti_20}.
The same mathematical framework was extended to scalar flux modelling, 
both for heat transfer~\citep{so_04a,younis_05,wang_07} and for buoyancy~\citep{so_04b}.
Similar to the original tensor polynomial expansion of~\citet{pope_75}, the turbulent heat flux is expanded as
\begin{equation}
\widetilde{\mathbf{q}} = \sum_{n=1}^{N}\alpha^{(n)}\mathbf{V}^{(n)},\quad N=1, \dots, 10
\label{eq:geneddy}
\end{equation}
where the coefficients $\alpha^{(n)}$ are scalar functions and $\mathbf{V}^{(n)}$ is a set of ten vector bases,
\begin{align}
\begin{aligned}
\mathbf{V}^{(1)}&= \boldsymbol\nabla\Theta \\
\mathbf{V}^{(2)}&= \mathbf{a}\boldsymbol\nabla\Theta\\
\mathbf{V}^{(3)}&= \mathbf{S}\boldsymbol\nabla\Theta\\ 
\mathbf{V}^{(4)}&= \mathbf{a}\mathbf{a}\boldsymbol\nabla{\Theta} \\
\mathbf{V}^{(5)}&= \mathbf{S}\mathbf{S}\boldsymbol\nabla{\Theta}\\ 
\end{aligned}
&&
\begin{aligned}
\mathbf{V}^{(6)} &= \mathbf{\Omega}\boldsymbol\nabla\Theta\\ 
\mathbf{V}^{(7)} &= \mathbf{\Omega}\mathbf{\Omega}\boldsymbol\nabla\Theta\\ 
\mathbf{V}^{(8)} &= \left(\mathbf{S}\mathbf{\Omega}+\mathbf{\Omega}\mathbf{S}\right)\boldsymbol\nabla\Theta\\
\mathbf{V}^{(9)} &= \left(\mathbf{a}\mathbf{S}+\mathbf{S}\mathbf{a}\right)\boldsymbol\nabla\Theta\\
\mathbf{V}^{(10)} &= \left(\mathbf{\Omega}\mathbf{S}+\mathbf{S}\mathbf{\Omega}\right)\boldsymbol\nabla\Theta, 
\end{aligned}
\label{eq:bases}
\end{align}
where $\boldsymbol{S} = 1/2 \left( \nabla \boldsymbol{U} + \nabla \boldsymbol{U}^T \right)$,
$\boldsymbol{\Omega} = 1/2 \left( \nabla \boldsymbol{U} - \nabla \boldsymbol{U}^T \right)$,
are the mean strain-rate and rotation-rate tensors, respectively.
As for the Reynolds stress tensor, in most practical flow cases the ten bases are not independent, and exact representation of
the turbulent heat flux ($\widetilde{\boldsymbol{q}}=\boldsymbol{q}$) is recovered for $N=3$.
In order to assess the accuracy of different truncations of the polynomial expansion~\eqref{eq:geneddy}
we compute the coefficients $\alpha^{(n)}$ following the approach of~\citet{jongen_98,modesti_20}, namely
we take the scalar product of~\eqref{eq:geneddy} with each vector basis $\mathbf{V}^{(m)}$, thus obtaining 
a linear system of $N$ equations for the unknown coefficients $\alpha^{(n)}$,
\begin{equation}
\mathbf{q}\cdot\mathbf{V}^{(m)} = \sum_{n=1}^{N}\alpha^{(n)}\mathbf{V}^{(n)}\cdot\mathbf{V}^{(m)},\quad m=1,\dots, N.
\label{eq:coeff}
\end{equation}
For $N=1$ this approach yields the standard definition of linear eddy diffusivity,
\begin{equation}
 \alpha_t=-\alpha^{(1)}=-\frac{\mathbf{q}\cdot\boldsymbol\nabla\Theta}{\boldsymbol\nabla\Theta\cdot\boldsymbol\nabla\Theta},
\label{eq:alphatdns}
\end{equation}
in analogy with the definition of linear eddy viscosity,
\begin{equation}
\nu_t=-\frac{\left\{\mathbf{a}\mathbf{S}\right\}}{2\left\{\mathbf{S}\mathbf{S}\right\}},
\label{eq:nutdns}
\end{equation}
where $\left\{\mathbf{AB}\right\} = A_{ik}B_{ki}$.
In the more general case one can solve equation~\eqref{eq:coeff} to find the coefficients $\alpha^{(n)}$ for different $1\leq N \leq 3$,
and then evaluate the modelled heat flux vector $\widetilde{\mathbf{q}}$ from~\eqref{eq:geneddy}.

\begin{figure}
 \begin{center}
  \includegraphics[scale=1.0]{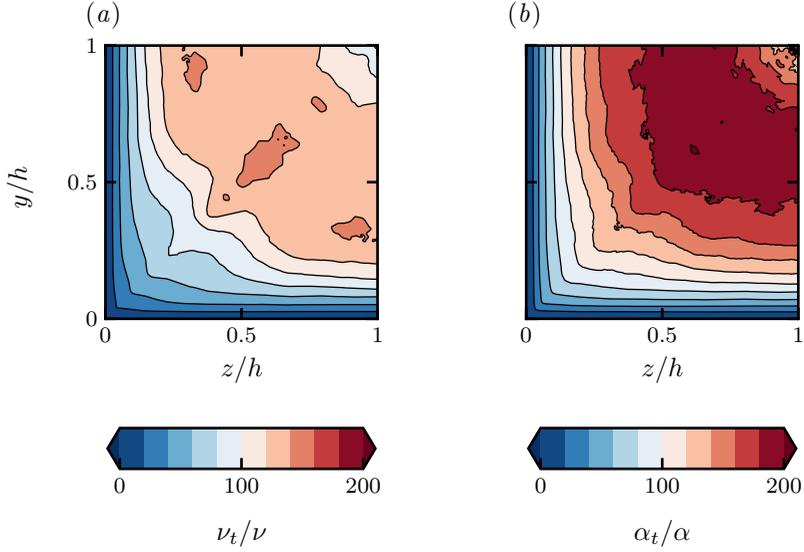}
	 \caption{Estimated fields of linear eddy viscosity $\nu_t$ as after equation~\eqref{eq:nutdns} 
        (\textit{a}), and linear eddy diffusivity $\alpha_t$ as after equation~\eqref{eq:alphatdns} (\textit{b}),
	          for flow case F ($\Rey_\tau=2000$). Only one quadrant of the duct is shown.
           }
  \label{fig:nut2d}
 \end{center}
\end{figure}

In order to assess the validity of analogy between turbulent momentum and heat transport,
in figure~\ref{fig:nut2d} we compare the linear eddy viscosity with the linear eddy diffusivity 
estimated from DNS flow case F.
Visual scrutiny suggests that the assumption $\Pran_t\approx 1$ is not valid
everywhere, and in fact contours of $\nu_t$ and $\alpha_t$ do not match. Furthermore, the eddy diffusivity
seems to be more affected by the presence of secondary flows as its iso-lines more markedly 
protrude towards the corner and tend to be more parallel to the walls as compared to $\nu_t$.

\begin{figure}
 \begin{center}
  \includegraphics[scale=1]{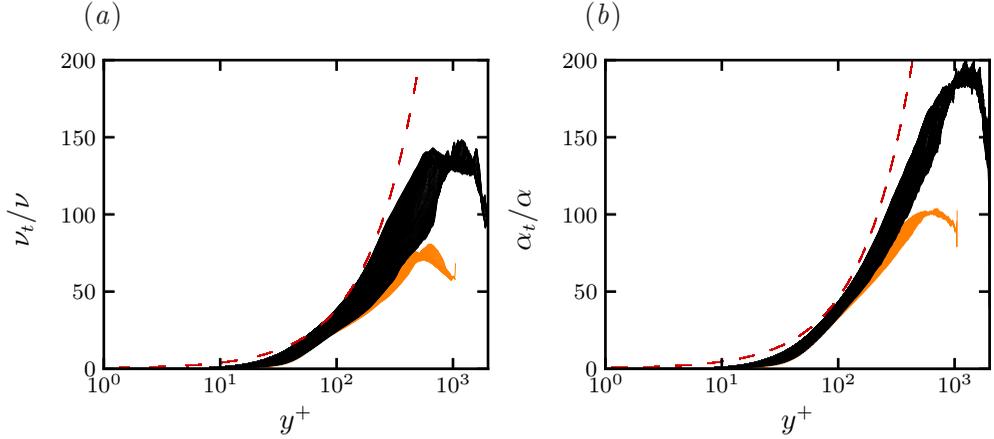}
	 \caption{Wall-normal distributions of linear eddy viscosity $\nu_t$ (\textit{a}) and linear eddy diffusivity $\alpha_t$ (\textit{b}) at all $z$ locations, 
	 up to the corner bisector ($y=z$, see figure~\ref{fig:phimean}) for flow cases E (orange lines) and F (black lines). The red dashed lines indicate
	 prediction of the mixing length hypothesis $\kappa y^+$ (\textit{a}) and $\kappa_\theta y^+$ (\textit{b}).}
  \label{fig:nut_bundles}
 \end{center}
\end{figure}

The spatial distribution of $\nu_t$ and $\alpha_t$ suggests that the value of the two scalars is controlled by
the closest wall, and therefore in figure~\ref{fig:nut_bundles} we report bundles of $\nu_t$ and $\alpha_t$ profiles up to the corner bisector.
Besides the obvious increase of $\nu_t$ and $\alpha_t$ with the Reynolds number, the figure shows
that the bundles of $\alpha_t$ have more limited scatter than $\nu_t$, confirming that the 
simple picture in which square duct flow is regarded as the superposition of two independent walls 
is more accurate for temperature than for momentum transport.
This effect may be interpreted as due to the additional effective diffusivity brought about by 
the non-local action of the pressure gradient term in the streamwise momentum equation, 
which as previously noted tends to smoothen the velocity field.
Likewise, it appears that it has also the effect of enhancing interactions between neighboring walls.
The red dashed lines in figure~\ref{fig:nut_bundles} show the result
of the mixing length hypothesis, namely $\nu_t/\nu\approx \kappa y^+$ and $\alpha_t/\alpha\approx \kappa_\theta y^+$,
which reveals to be only partially accurate in the overlap layer, especially for the eddy diffusivity.

\begin{figure}
 \begin{center}
  \includegraphics[scale=1]{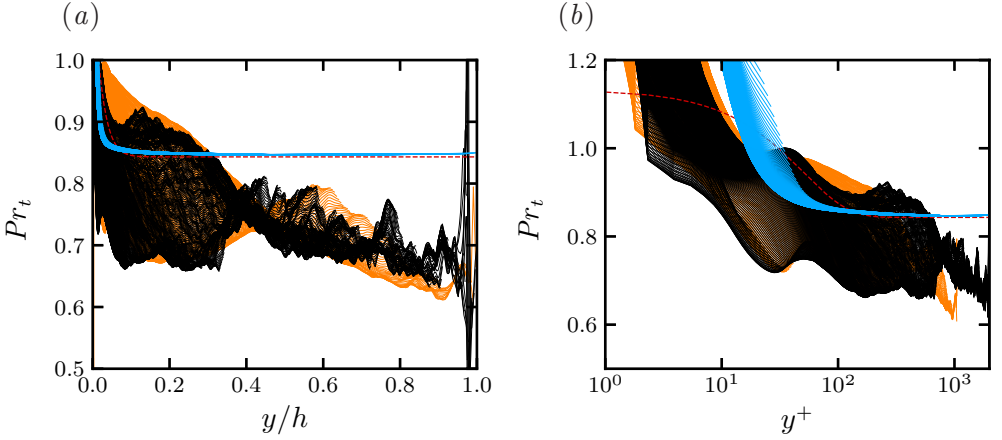}
	 \caption{Wall-normal distributions of turbulent Prandtl number $\Pran=\nu_t/\alpha_t$, in outer coordinates (\textit{a}) and in inner coordinates (\textit{b}). Profiles are plotted at all $z$ locations, up to the corner bisector ($y=z$, see figure~\ref{fig:phimean}). 
	 Flow cases E ($\Rey_\tau=1000$, orange) and F ($\Rey_\tau=2000$, black) are compared to equation~\eqref{eq:pran_cebeci} of~\citet[red dashed]{cebeci_73} and equation~\eqref{eq:pran_kays} of~\citet[blue]{kays_93}.}
  \label{fig:prt}
 \end{center}
\end{figure}

Figure~\ref{fig:prt} shows bundles of the turbulent Prandtl number~\eqref{eq:prt} for flow case E and F, as a function of the wall distance.
We also report predictions of available refined models for the turbulent Prandtl number, including that developed by~\citet{cebeci_73},
\begin{equation}
	\Pran_t = \frac{k}{\kappa_\theta}\left(\frac{1-\exp{(-y^+/A)}}{1-\exp{(-y^+/B)}}\right),\quad B=\frac{1}{\Pran^{(1/2)}}\sum_{i=1}^5C_i\left(\log_{10}\Pran\right)^{(i-1)},
	\label{eq:pran_cebeci}
\end{equation}
where $A=26$, $C_1=34.96$, $C_2=28.79$, $C_3=33.95$, $C_4=6.3$, $C_5=-1.186$, and that developed by~\citet{kays_93},
\begin{equation}
	\Pran_t=\left[\frac{1}{2\Pran_{t,b}} + \frac{C\Pen_t}{\sqrt{\Pran_{t,b}}} - (C\Pen_t)^2\left(1-\exp{\left(-\frac{1}{C\Pen_t\sqrt{\Pran_{t,b}}}\right)}\right)\right]^{-1},
	\label{eq:pran_kays}
\end{equation}
where $\Pen_t=\Pran \, \nu_t/\nu$ is the turbulent Peclet number, $\Pran_{t,b}=0.84$ is the assumed bulk turbulent Prandtl number (namely, away from walls), and $C=0.3$.
Figure~\ref{fig:prt} shows that $\Pran_t$ is far from universal across the wall-normal stations, and large scatter
of the bundles is observed near the wall, whereas slightly better behavior is observed at $y/h \gtrsim 0.5$.
We further note that in the outer region the turbulent Prandtl number attains values around 0.7, which are significantly 
lower than in plane channel flow~\citep{pirozzoli_16}.
The engineering correlations do not perform well. The model of~\citet{cebeci_73} largely overpredicts $\Pran_t$
in the duct core, and it does not reproduce the correct trend close to the wall. 
Also the model of~\citet{kays_93} overpredicts the turbulent Prandtl number in the bulk flow, 
and it does not account accurately for spatial non-uniformities.
 
We further assess the validity of the nonlinear eddy diffusivity model~\eqref{eq:geneddy} for different truncations of the vector polynomial basis.
Here we only focus the $q_1$ and $q_2$ components of the heat flux vector because $q_2(y,z)=q_3(z,y)$ after geometrical symmetry.
Several nonlinear constitutive relations can be obtained with different combinations of the vector integrity bases~\eqref{eq:bases},
providing different accuracy for the modelled heat flux vector.
In order to quantify the error with respect to DNS, we use the averaged correlation coefficient between $q_{i}$ and $\widetilde{q}_{i}$,

\begin{equation}
\widetilde{C}_{i} = \frac{\langle q_{i}\widetilde{q}_{i}\rangle-\langle q_{i}\rangle\langle \widetilde{q}_{i}\rangle }{\left[\left(\langle q_{i}^2\rangle-\langle q_{i}\rangle^2\right)\left(\langle \widetilde{q}_{i}^2\rangle-\langle \widetilde{q}_{i}\rangle^2\right)\right]^{1/2}},
\label{eq:err}
\end{equation}
where the angle brackets denote average over the duct cross section.
By construction $\widetilde{C}_i\in[-1,1]$, where $\widetilde{C}_i=1$ indicates perfect correlation, $\widetilde{C}_i=-1$ negative correlation.

\begin{table*}
 \begin{center}
\begin{tabular}{lrrr}
\hline
 N &  $V^{(n)}$  &   $\widetilde{C}_{1}$ &   $\widetilde{C}_{2}$  \\
\hline
1 & 1 & - &  0.997  \\
\hline
\multirow{4}{*}{2} & 1,2 & 0.538 &  0.973  \\
  &{\bf 1,3} & {\bf 1.000} & {\bf 0.988} \\
	&1,4 & 0.871 & 0.997 \\
  &1,5 & 0.896 & 0.801\\
\hline
\multirow{6}{*}{3}  &1,2,3& 0.999 & 1.000  \\
  &1,2,4& 0.538 & 0.973   \\
  &1,2,5& 0.988 & 1.000    \\
  &1,3,4& 1.000 & 0.988 \\
  &{\bf 1,3,5}& {\bf 1.000} & {\bf 1.000}  \\
  &1,4,5& 0.892 & 0.801  \\
\hline
\end{tabular}
 \caption{
          Correlation coefficient $\widetilde{C}_{i}$ \eqref{eq:err} between the turbulent heat flux of DNS $q_{i}$ and
          modelled turbulent heat flux $\widetilde{q}_{i}$, equation~\eqref{eq:geneddy},
	  for flow case F and
          for different combinations of vector bases. The optimal subset for each $N$ is highlighted in boldface.}
 \label{tab:err_bases_subset}
 \end{center}
\end{table*}
Table~\ref{tab:err_bases_subset} shows the correlation coefficients $\widetilde{C}_{1}$ and 
$\widetilde{C}_{2}$ for flow case F, where we consider all possible bases combinations
that can be obtained using the first five bases, and up to $N=3$.
The analysis shows that it is possible to find optimal bases subsets providing the maximum accuracy
for a given $N$ (boldface values in table~\ref{tab:err_bases_subset}).
The linear eddy diffusivity hypothesis ($N=1$) is able to accurately predict the turbulent heat flux component $\widetilde{q}_2$,
whereas $\widetilde{q}_1=0$ by construction. In the case of $N=2$ we find that the bases combination $\mathbf{V}^{(1)},\mathbf{V}^{(3)}$ is the one 
which brings the highest accuracy on $\widetilde{q}_1$ and also a very good prediction of $\widetilde{q}_2$, although
slightly lower than the linear eddy diffusivity hypothesis. 
The choice $\mathbf{V}^{(1)},\mathbf{V}^{(4)}$ leads to the same accuracy on $\widetilde{q}_2$ as for the linear case,
but it increases the error on $\widetilde{q}_1$. We further note that $\mathbf{V}^{(4)}$
contains the anisotropic Reynolds stress tensor, and therefore it is not a practical choice from the modelling point of view.
Finally, we find that using three vector bases is sufficient to recover the exact representation of the turbulent heat flux vector
if the subset $\mathbf{V}^{(1)},\mathbf{V}^{(3)},\mathbf{V}^{(5)}$ is used. Notably, 
these vector bases do not contain the anisotropic Reynolds stress tensor, whose presence would certainly lead to 
a much lower accuracy in a practical scenario, as it also requires modelling.

\begin{figure}
 \begin{center}
  \includegraphics[scale=1.0]{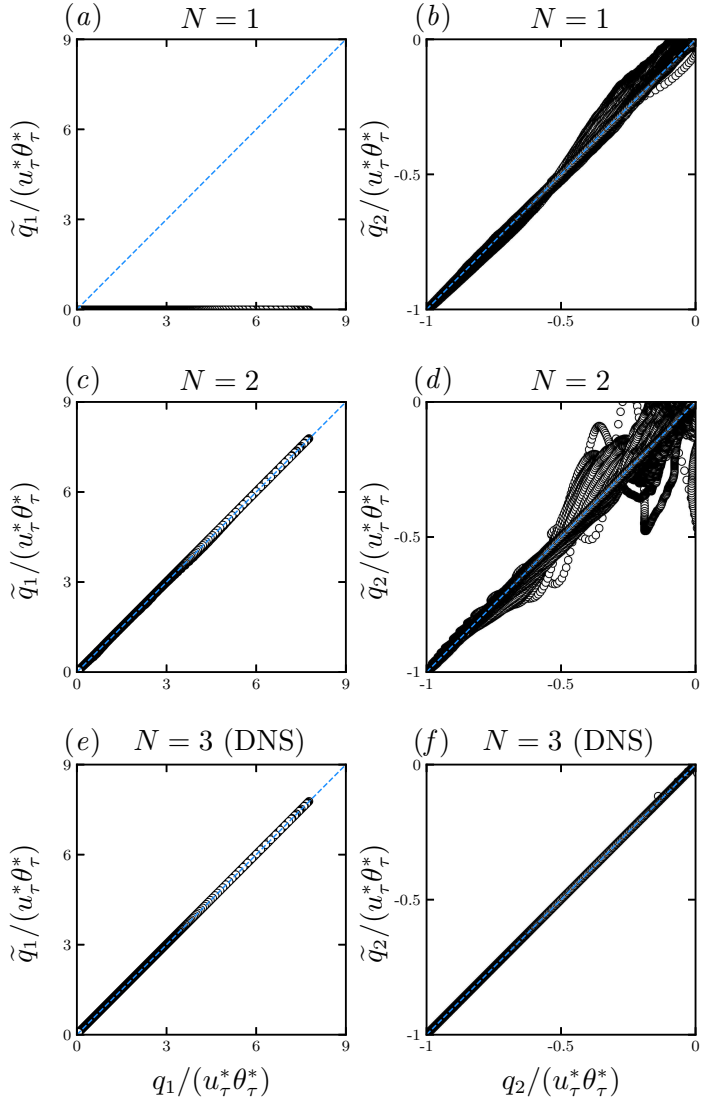}
	 \caption{Scatter plots of the modelled turbulent heat flux vector components ($\widetilde{q}_1$, $\widetilde{q}_2$),
	 versus those obtained from DNS ($q_1$, $q_2$) for flow case F, for increasing number of vector bases: $N=1$ (\textit{a}, \textit{b}),
	 $N=2$, $\mathbf{V}^{(1)},\mathbf{V}^{(3)}$ (\textit{c}, \textit{d}) and $N=3$, $\mathbf{V}^{(1)},\mathbf{V}^{(3)},\mathbf{V}^{(5)}$ (\textit{e}, \textit{f}). 
         The dashed diagonal line indicates ideal model performance.}
  \label{fig:scatter_q}
 \end{center}
\end{figure}
 
Figure~\ref{fig:scatter_q} shows scatter plots of the modelled turbulent heat flux components versus those obtained from DNS, 
in which the axis bisector indicates ideal model behavior. Here we limit ourself to plotting the results for the linear eddy diffusivity hypothesis
and for the optimal bases subsets highlighted in table~\ref{tab:err_bases_subset}.
The linear eddy diffusivity hypothesis returns accurate prediction of $\widetilde{q}_2$, but it is not able to predict $\widetilde{q}_1$, 
which reminds us of difficulties of linear eddy-viscosity models in predicting the normal Reynolds stress components. 
Using two vector bases improves the prediction of $\widetilde{q}_1$, but it yields less accurate approximation
of $\widetilde{q}_2$, which would negatively impact prediction of the turbulent heat flux. Similar considerations apply to
the anisotropic Reynolds stress tensor when two tensor bases are used~\citep{modesti_20}, although less evident than here.
Finally, exact representation of the heat flux vector is recovered for $N=3$.
 
\begin{figure}
 \begin{center}
  \includegraphics[scale=1.0]{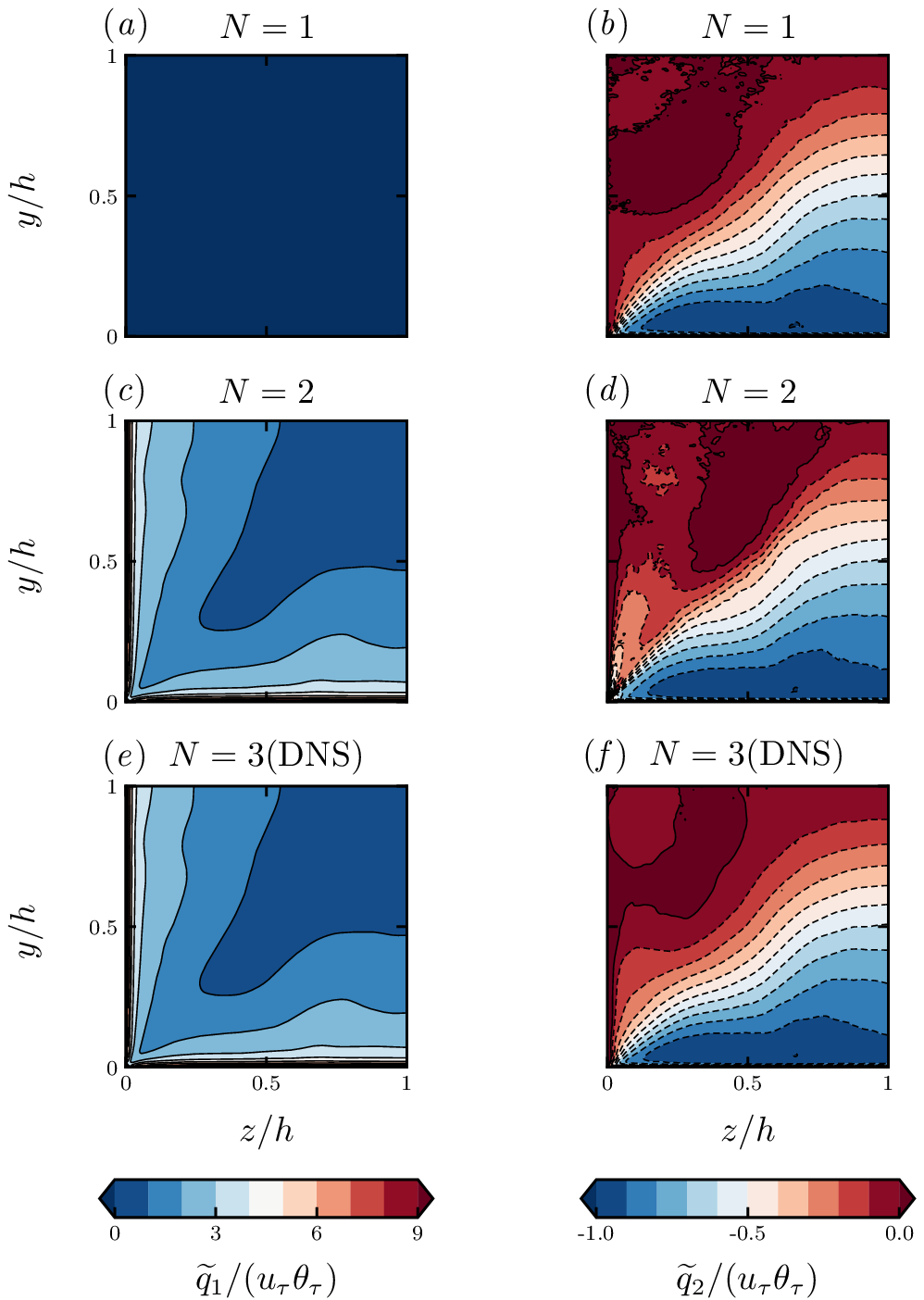}
	 \caption{Modelled turbulent heat flux vector components ($\widetilde{q}_1$, $\widetilde{q}_2$) in a duct quadrant
	 for flow case F and
	 increasing number of vector bases: $N=1$ (\textit{a}, \textit{b}),
	 $N=2$ $\mathbf{V}^{(1)},\mathbf{V}^{(3)}$ (\textit{c}, \textit{d}), $N=3$ $\mathbf{V}^{(1)},\mathbf{V}^{(3)},\mathbf{V}^{(5)}$ (\textit{e}, \textit{f}).
           }
  \label{fig:qmod2d}
 \end{center}
\end{figure}
 
Additional information on the accuracy of nonlinear eddy diffusivity models can be extracted from the
spatial distribution of $\widetilde{q}_1$ and $\widetilde{q}_2$ for increasing number of tensor bases, figure~\ref{fig:qmod2d}.
When linear representation is used $\widetilde{q}_2$ closely resembles the DNS data, whereas $\widetilde{q}_1=0$.
Using two vector bases improves the prediction of $\widetilde{q}_1$, which becomes indistinguishable from DNS data, whereas
the modeling accuracy 
of $\widetilde{q}_2$ substantially decreases, especially close to the side wall.
For $N=3$ the exact heat flux vector is recovered.
This analysis reveals that the linear eddy-diffusivity hypothesis has the potential to accurately predict $\langle{v'\theta'}\rangle$,
and therefore the overall heat flux. However, this good performance is likely to be hampered by the use of a constant turbulent Prandtl number,
which is far from being accurate.
Slightly better prediction of $\widetilde{q}_2$ can be achieved using a nonlinear eddy diffusivity with at least three vector bases, 
but improvement in the prediction $\widetilde{q}_2$ might not justify the additional modelling complexity.

\section{Conclusions}

We have carried out DNS of square duct flow at the unprecedented Reynolds number of $\Rey_\tau\approx 2000$, wherein Navier--Stokes equations
have been augmented with the transport of a passive scalar representing the temperature field. This configuration corresponds
to the case of forced convection in which hot fluid is pumped through the duct and cooled at the walls.
The choice of using a unity Prandtl number and homogeneous cold walls allows us a direct comparison between heat and momentum transport.
In this respect, we find that the streamwise velocity and temperature fields are very similar, both instantaneously and on average. 
Close to the walls $U$ and $\Theta$ are nearly identical because the fluid viscosity and thermal diffusivity dominate, and $\Pran=1$,
whereas in the overlap layer they both follow a logarithmic law, although with a different slope.
At low Reynolds number the correlation coefficient between $u$ and $\theta$ is less than one at the duct core,
but it but it approaches the unit value as the flow becomes well mixed, restoring the similarity inherent in the underlying equations.

The Nusselt number of square duct flow is in excellent agreement with pipe flow data at matching Reynolds number, which
support the validity of the hydraulic diameter concept.
We explain this good match by using a simple model in which the duct flow can be regarded as the superposition
of four concurrent walls, where the flow is controlled by the closest one.
This simple cartoon is well supported by the mean temperature profiles in a duct octant, which follow with good accuracy the canonical law-of-the-wall.

We have derived a generalized version of the popular FIK identity for the heat flux, which confirms the common idea that secondary flows
increase the heat transfer by mixing hot fluid at the duct core with cold fluid at the corners. 
We have also carried out numerical simulations whereby secondary flows have been artificially suppressed, which
also confirm minor effect of the secondary currents on the global heat transfer.
Although the secondary flows are able to bend the
isolines of the mean velocity and temperature, their global effect on heat transfer is rather small and can be estimated to be about $5\%$.

The DNS data have been used to establish the accuracy of linear and nonlinear eddy diffusivity closures for the turbulent heat flux.
The commonly accepted hypothesis of uniform turbulent Prandtl number is highly inaccurate in large part of the duct, and at the duct core $\Pran_t\approx0.7$,
which is much lower than what found in channel and pipe flow ($\Pran_t\approx0.85$).
We show that exact representation of the heat flux vector can be recovered using a vector polynomial integrity basis expansion with at least three bases,
although this seems impractical from a modelling point of view.
The linear eddy diffusivity formulation is able to accurately predict the wall-normal turbulent heat flux,
and therefore it can be a much more reliable alternative to the turbulent Prandtl number, 
provided that an accurate transport equation for the eddy conductivity is available.\\

\noindent
DNS data are available at http://doi.org/10.4121/19221657 and at\\ http://newton.dima.uniroma1.it/database/ \\

\noindent
Declaration of Interests. The authors report no conflict of interest.\\

\noindent
{\bf Acknowledgements}\\
We acknowledge that the results reported in this paper have been achieved using 
the PRACE Research Infrastructure resource MARCONI based at CINECA, Casalecchio di Reno, Italy,
and the DECI Research Infrastructure Beskow at PDC, Stockholm, Sweden.

\addcontentsline{toc}{chapter}{Bibliography}
\bibliographystyle{plainnat}
\bibliography{references} %

\end{document}